\documentclass[journal]{new-aiaa}
\usepackage[utf8]{inputenc}
\usepackage{graphicx}
\usepackage{amsmath}
\usepackage[version=4]{mhchem}
\usepackage{siunitx}
\usepackage{longtable,tabularx,color}
 \usepackage{subfigure}% subcaptions for subfigures
 \usepackage{subfigmat}% matrices of similar subfigures, aka small mulitples
\usepackage{caption}
\usepackage[section]{placeins}

\title{A Study of Side Jet Interactions over a Hypersonic Cone Flow using Kinetic Methods}

\author{Irmak T. ~Karpuzcu\footnote{Graduate Student, Department of Aerospace Engineering, AIAA Student Member} and Deborah A.~Levin\footnote{Professor, Department of Aerospace Engineering, AIAA Fellow.}}
\affil{University of Illinois, Urbana-Champaign, IL 61801, USA}

\begin{document}

\maketitle
%\tableofcontents

\let\footnote\relax\footnotetext{Partially presented at AIAA Aviation 2022 Forum on June 27-July 1, 2022 at Chicago, IL and Virtual with the paper number AIAA 2022-3987 and at AIAA SCITECH 2023 Forum on 23-27 January 2023 at National Harbor, MD and Online with the paper number 2023-0857.}

\begin{abstract}
The interaction between the hypersonic flow over a cone and a side jet  creates complex multiple recirculation regions as well as separation, reattachment and bow shocks known as Edney-Type IV shock-shock interactions.  The interactions between these multi-scale structures are shown to be well captured with the direct simulation Monte Carlo (DSMC) method, a particle-kinetic approach.  This computational technique was used because even though the free stream conditions were continuum-like, sufficiently strong gradients in the shock and jet expansion were observed to cause continuum breakdown.  The time-accurate DSMC approach has been shown in previous work to capture the multi-scale structures and unsteadiness in the shock structure that occurs in such flows, as was also observed in this work.  These sources of unsteadiness in the shock system are shown to locally affect surface parameters such as heat transfer and skin friction as well as predict a low frequency unsteadiness consistent with earlier values in the literature. 
\end{abstract}

\section*{Nomenclature}

%\noindent(Nomenclature entries should have the units identified)

{\renewcommand\arraystretch{1.0}
\noindent\begin{longtable*}{@{}l @{\quad=\quad} l@{}}
$A_b$ & base area of the cone, $m^2$\\
$c_f$& skin friction coefficient \\
$c_h$& heat transfer coefficient \\
$D_{b}$& base diameter of the cone, $m$ \\
$f$& frequency, $Hz$ \\
$F_D$ & drag force acting on cone, $N$ \\
$F_N$ & normal component of the resultant pressure force acting on the cone, $N$ \\
$FNUM$ & number of real particles each simulation particle represents \\
$h$ & penetration height of the jet gas, $m$ \\
$L$ & length of the cone, $m$ \\
$M$   & Mach number \\
$p$ & pressure, $Pa$\\
$P$ & continuum breakdown parameter\\
$Re$ & Reynolds number\\
$r_b$ & radius of the base of the cone, $m$\\
$R_j$ & gas constant of the jet gas, $\frac{J}{{kgK}}$\\
$St$ & Strouhal number\\
$T_{0j}$& stagnation temperature of the jet gas, $K$\\
$U_\infty$  & free stream velocity, $m/s$\\
$_j$ & properties of the jet gas\\
$\Delta t$ & time step or time difference, $s$ \\
$\gamma $ &ratio of the specific heats \\
$\lambda$   & mean free path, $m$ \\
$\widetilde m$  & normalized mass flow rate of the jet gas\\
$\dot m$  & mass flow rate of the jet gas, $kg/s$ \\
$\rho$  & mass density, $kg/m^3$ \\
$\theta_{SL}$  & angle between the shear layer and the cone surface, degrees \\
$\theta_{BS}$  & angle between the bow shock and the cone surface, degrees \\
$\tau_{flow}$ & flow time, $s$
\end{longtable*}}

\section{Introduction}\label{sec:Intro}
Because side jets are one of the widely used control mechanisms for maneuvering and re-entry for high Mach number vehicles~\cite{CasselReview2003}, the flow structure caused by the jet - shock wave interactions and the possible effects of these flow structures on surface properties such as drag forces and heat transfer has been the focus of many studies~\cite{cassel1973,roger1999,mitchell1964}. Understanding these phenomenon, which are known as aerodynamic interference effects, will aid in the design of control mechanisms of such vehicles. The specific flow structure and the effect of the jet depends on the flight configuration and the geometry of the vehicle and control surfaces as well as the flight regime. The effect of a side jet will cause the flow structure to be complex due to separation and separation related shock waves upstream of the jet and circulation and reattachment of the flow downstream of the jet. The overall characteristics of these flow fields depend on whether the free stream flow is supersonic or hypersonic, whether the incoming boundary layer is laminar or turbulent, and several other parameters such as penetration height of the jet, momentum ratio of the jet to the free stream etc.

\begin{figure}[htbp]
\begin{center}
{\includegraphics[trim=0cm 0cm 0cm 0cm,clip,width=0.66\linewidth]{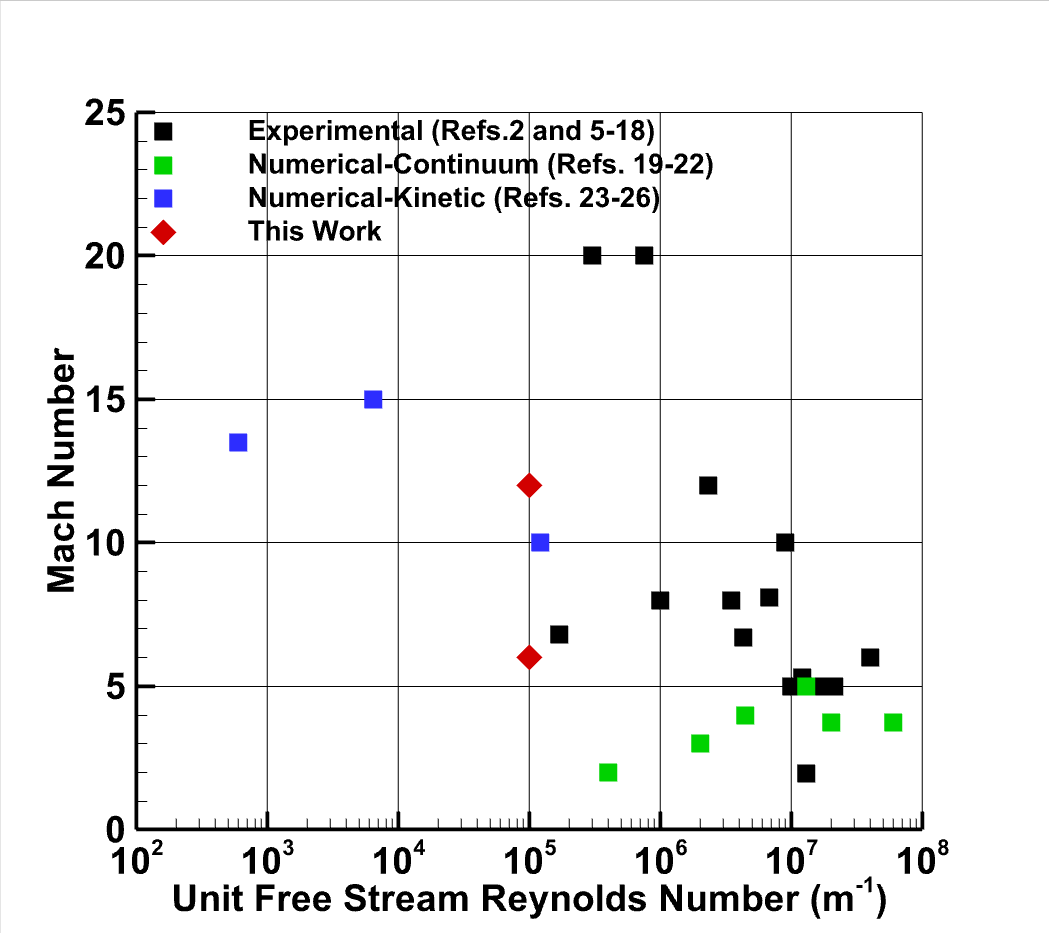}}
\end{center}
\caption{Free stream Mach and unit Reynolds numbers of the previous related studies.}
\label{fig:LitReview}
\end{figure}

A review of the literature shows that most previous studies of side jets, or traverse jets, are for lower altitudes of unit Re numbers of $\sim10^{7}$ $m^{-1}$. Figure~\ref{fig:LitReview} shows the unit Reynolds number versus free stream Mach numbers for some of these studies. Most of the earlier work obtained measurements on several different  configurations of the jet and test article, ~\cite{Mair1952,Maull1960,Barber1966,Kaufman1967,Shreeve1972,cassel1973,Rosenbaum1975,Shang1989,Yeneriz1989,Roberts1998,Kurita2000,Gulhan2008,Taguchi2015,Beresh2018,Erdem2021}, however,  a number of numerical studies using continuum methods~\cite{Lee2006,Chakraborty2010,Miller2018,Sinha2021} were also performed. The studies shown in Fig.~\ref{fig:LitReview} included both laminar and turbulent boundary layers, different pressure ratios of jet to free stream, and several jet outlet conditions.  In general, it was observed that the Mach number of the incoming boundary layer and the pressure ratio of the jet to free stream are the two important parameters that affect surface parameters such as drag force and heat transfer.  There are also a few studies that use kinetic methods but almost all of them are for higher altitudes~\cite{LevinSideJet2002,ozawasidejet2007,li2007,GlassFP}. In these earlier works, the main focus was the expansion of a jet to a rarefied environment and in some cases chemical reactions inside the jet expansion were also considered. However with the advances in computational algorithms and hardware we can use  the  direct simulation Monte Carlo (DSMC)~\cite{Bird} for continuum-like free stream number densities, as indicated by the red diamonds in Fig.~\ref{fig:LitReview}. 

The most important aspect of modeling flows with side jets is the jet expansion since it affects all the flowfield characteristics in the interaction region~\cite{GlassFP,KARAGOZIAN2010}.  The jet flow causes a blockage to the incoming flow which then results in separation of the boundary layer upstream of the jet. The extent of this separation and the height of the jet gas penetration depends strongly on the jet expansion characteristics. For example, outlet speed of the jet was shown in Ref.~\cite{CasselReview2003} to change the expansion shape and structures of the interaction zone.  Moreover, the expansion of the jet occurs in a very small spatial region compared to the simulation domain suggesting that there will be high gradients occurring in this region. Since the  incoming flow is still supersonic even after the leading edge shock in hypersonic flows for most cases, the presence of the jet causes several shocks and strong shear layers, regardless of the free stream Reynolds number. The presence of these high gradient regions will be shown in this work to cause continuum breakdown. Therefore,  there is a strong need to use kinetic approaches, such as the DSMC method, to simulate such flows.  

DSMC is a well established time accurate method to model flows with high gradients since it can capture shocks and shear layers without problems due to numerical viscosity.  Our previous work with DSMC simulations included several geometries such as double wedge and double cone geometries and compression ramps that all create separation bubbles and several shocks and shear layers. We have characterized the unsteadiness in the shock system of a double cone configuration which is caused by  shock-boundary layer and shock-shock interactions~\cite{tumukluPoF,tumukluPoF2}. We have also discovered instabilities occurring in the shock layer which are in-sync with instabilities in the shear layer for the double wedge flow in Ref.~\cite{sawant_theofilis_levin_2022}.   As is well known,  these types of shock-shock and shock-boundary layer interactions can cause unsteadiness in the entire shock system~\cite{edney1968}, however, the specific importance of the shock system versus the boundary layer was previously unknown.  The oblique shock-side jet interaction creates a rich set of shock-shock interactions in a similar manner.  Yet to date, the main effort in the literature in modeling side jet interactions has been to characterize steady state surface parameters, and only a small number of studies studied unsteadiness of the flowfield and effect of the unsteadiness on surface parameters~\cite{Mair1952,Maull1960,Rosenbaum1975,Kurita2000}. Similar to our previous modeling of laminar shock-shock and shock-boundary layer interactions, we analyze the unsteadiness of gas-surface parameters in side-jet interactions using the
 DSMC, an inherently time-accurate method.

Our earlier previous  modeling of side jet interactions in a hypersonic flow we compared the effect of 
different fundamental geometries~\cite{KarpuzcuSciTech2022}, {\em e.g.},  flat plate, wedge and cone configurations, 
for a free stream Mach number of 6.   We found from these studies that the interaction region was the largest for the flat plate case, but,  the surface parameters changed the most for the cone geometry compared to the case with no side jet. Preliminary experiments have been conducted at the Purdue University's BAM6QT quiet tunnel~\cite{KarpuzcuSciTech2022,ChinskeSciTech2023} with a 7$^{\circ}$ half angle cone at zero angles of attack.  Qualitative agreement between our simulations and the experiments  in terms of shock structures and flow unsteadiness was observed, but, more detailed comparisons are still required.  
 In this work, we present flowfield and surface forces for the same cone geometry used at BAM6QT in both the preliminary side jet experiments~\cite{KarpuzcuSciTech2022, ChinskeSciTech2023} and transition studies~\cite{SchneiderReview2004,SchneiderHistory2019,kennedy_jewell_paredes_laurence_2022}.  In addition, we consider two different Mach numbers, one the same as the experiments at M=6 and the second  at M=12, to understand how the flow physics in the interaction region changes.  Our primary goals in this work are to investigate the overall flowfield characteristics, surface parameters, unsteadiness in the flowfield due to shock-shock and shock-boundary layer interactions and the effect of this unsteadiness on the local surfaced parameters.

The organization of the paper is as follows.  In Section.~\ref{sec:timeAveFlow} we discuss the specific cases in terms of their free stream and numerical DSMC parameters and present time averaged flowfields.  Following that, in Section.~\ref{sec:timeAveSurface} we  present the effect of the jet presence
on the steady state surface parameters. In Section.~\ref{sec:InstaFlow} we consider the instantaneous flowfield results and characterize the frequencies of the unsteadiness and its effects on surface parameters. Finally in Section~\ref{sec:Conc} we  summarize our work.

\section{Time Averaged Steady State Flowfield Results}\label{sec:timeAveFlow}

In this work, hypersonic flows over a cone with a 7$^{\circ}$ half angle were simulated using the Statistical Modeling In Low-density Environment (SMILE)~\cite{Ivanov-dsmctool} as the computational solver. A sonic jet was placed at the middle of the cone surface for both the Mach 6 and 12 flows with a working fluid of molecular nitrogen for both the free stream and jet gas. The cone has a base diameter and a length of $0.1$ and $0.4$~m, respectively and the jet has a diameter of $1$~mm.  In normalizing the spatial dimensions we use the cone length and we put the origin of the flowfield at the cone tip. Thus, the cone starts at $x/L=0.0$, ends at $x/L=1.0$, and the center of the jet is located at $x/L=0.5$, $y/L=0.0625$.  As mentioned above, 
the dimensions of the cone were selected to be the same as preliminary experiments conducted at BAM6QT at M=6~\cite{ChinskeSciTech2023}.  The free stream and jet flow properties are given in Table~\ref{tab:FreeStreamandJet} for a  unit free stream Reynolds number of $1\times10^{5}$ m$^{-1}$.  

In terms of numerical parameters, the total number of simulation particles was on the order of $500\times10^{6}$, with at least 10 particles in each collision cell locally for both cases with a time step of $\Delta t=5.0\times10^{-9}s$. The flow was simulated as an {\it axisymmetric}case with the x-axis as the symmetry axis and the surface of the cone was treated as an isothermal wall at $T_w=300K$. The computational domain had a length of $0.425$~m in the x-direction and a height of $0.15$~m in the y-direction. The time required for the flow to establish in the interaction region is 2.5 $ms$, which corresponds to 500,000 time steps. This time is determined by monitoring the numerical probes placed in the flowfield from the beginning of the simulation at time $t=0$. We start time sampling after this time instant and average for 100,000 time samples which correspond to 0.5 $ms$ to have adequate time averaged results. The computations were done on TACC's Stampede2 supercomputer. On this supercomputer, a run of 5$ms$ in physical time required 460,000 CPU-hours on Intel Xeon Phi 7250 ("Knights Landing") processors.

SMILE employs the majorant frequency scheme~\cite{majorant} for modeling collisions. To accurately model viscosity-temperature dependence, the Variable Hard Sphere (VHS)~\cite{Bird} model  was used with a nitrogen molecular diameter of $4.467$\AA{} and a viscosity index of $0.26$ at a reference temperature of $273K$. In addition, it was assumed that there were no chemical reactions, which is reasonable,  since the total temperature of the free stream and jet flows were lower than $1500~K$.

\begin{table}[htb!]
\caption{Free stream and jet flow parameters for the M=6 and M=12 case simulations.}
\label{tab:FreeStreamandJet}
\begin{center}
\scalebox{1.0}{
\begin{tabular}{ccccc}
\hline
\hline
  &M6&&M12&\\
\hline
Parameter  &Free Stream&Jet&Free Stream&Jet\\
\hline
Number Density(m$^{-3}$) & $ 1.00\times10^{22}$ &$ 1.65\times10^{23}$&$ 0.50\times10^{22}$&$ 8.25\times10^{23}$\\
Ma Number & 6.0 &1.0&12.0&1.0\\
Temperature  (K) & 50.0 &250.0&50.0&250.0\\
Pressure (Pa) & 6.9 &571.2&3.45&2847.4\\
Kn Number$^a$ & 0.07 &0.0067&0.15&0.0013\\
\hline
\hline
%\end{tabularx}
\multicolumn{5}{l}{$^a$ based on the reference length of the jet diameter, $d=1mm$}\\
\end{tabular}}
\end{center}
\end{table}

\begin{figure}[htbp]
\begin{center}
\subfigure[M6 case]{\label{fig:MaContourM6}\includegraphics[trim=0cm 0cm 0cm 9cm,clip,width=0.48\linewidth]{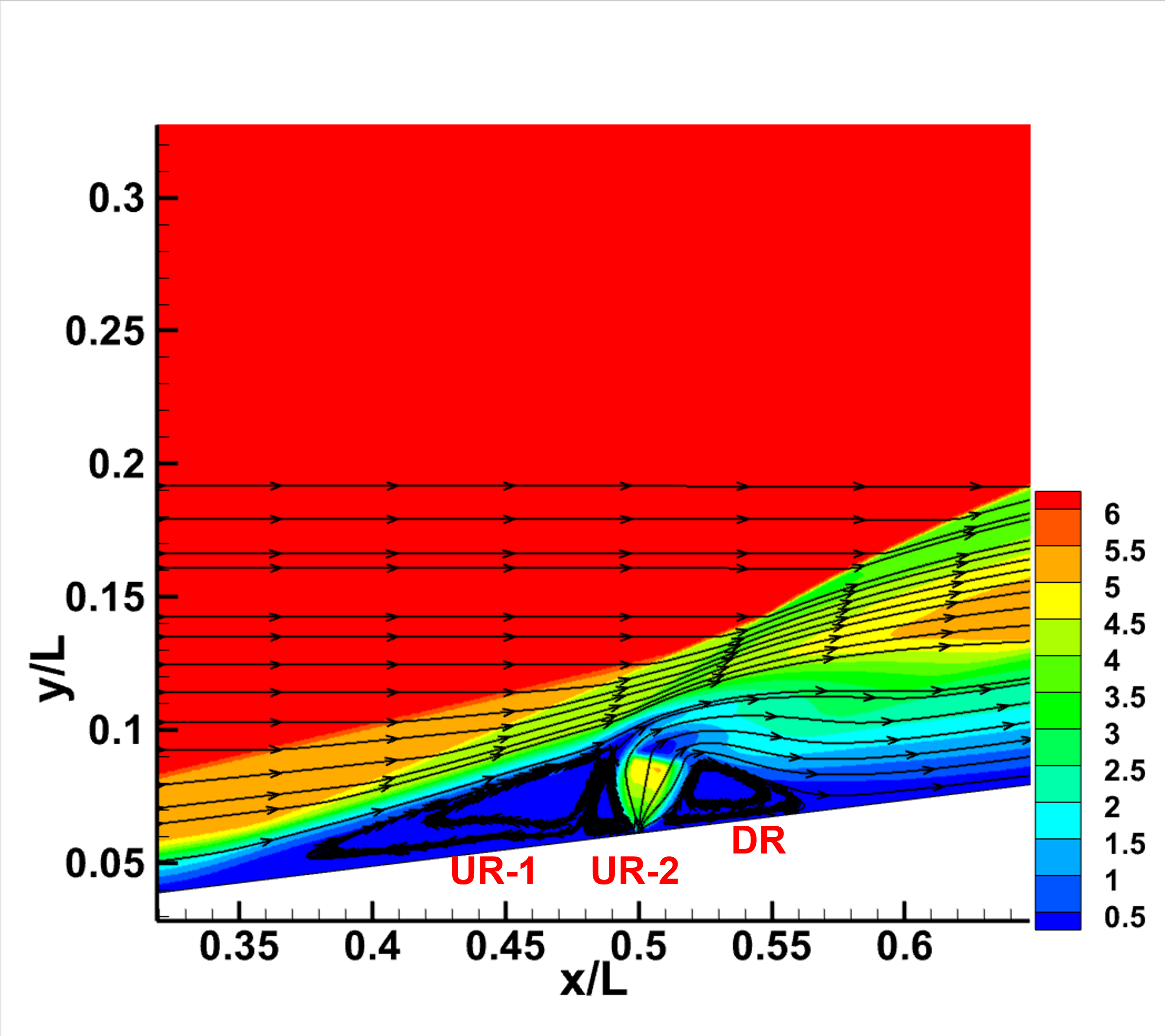}}
\subfigure[M12 case]{\label{fig:MaContourM12}\includegraphics[trim=0cm 0cm 0cm 9.3cm,clip,width=0.48\linewidth]{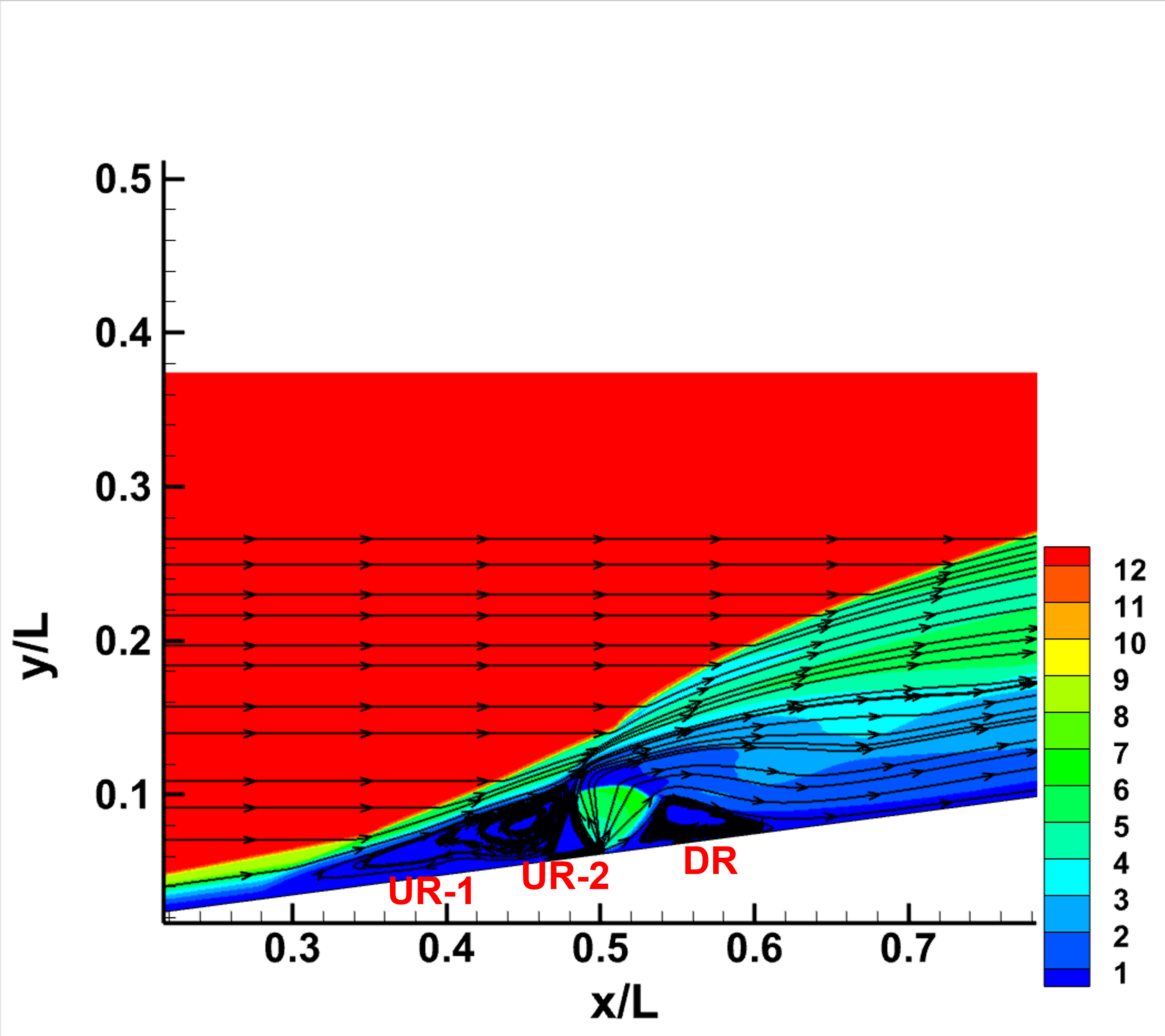}}
\subfigure[M6 case]{\label{fig:SchlierenM6}\includegraphics[trim=0.5cm 0.5cm 0.5cm 9cm,clip,width=0.48\linewidth]{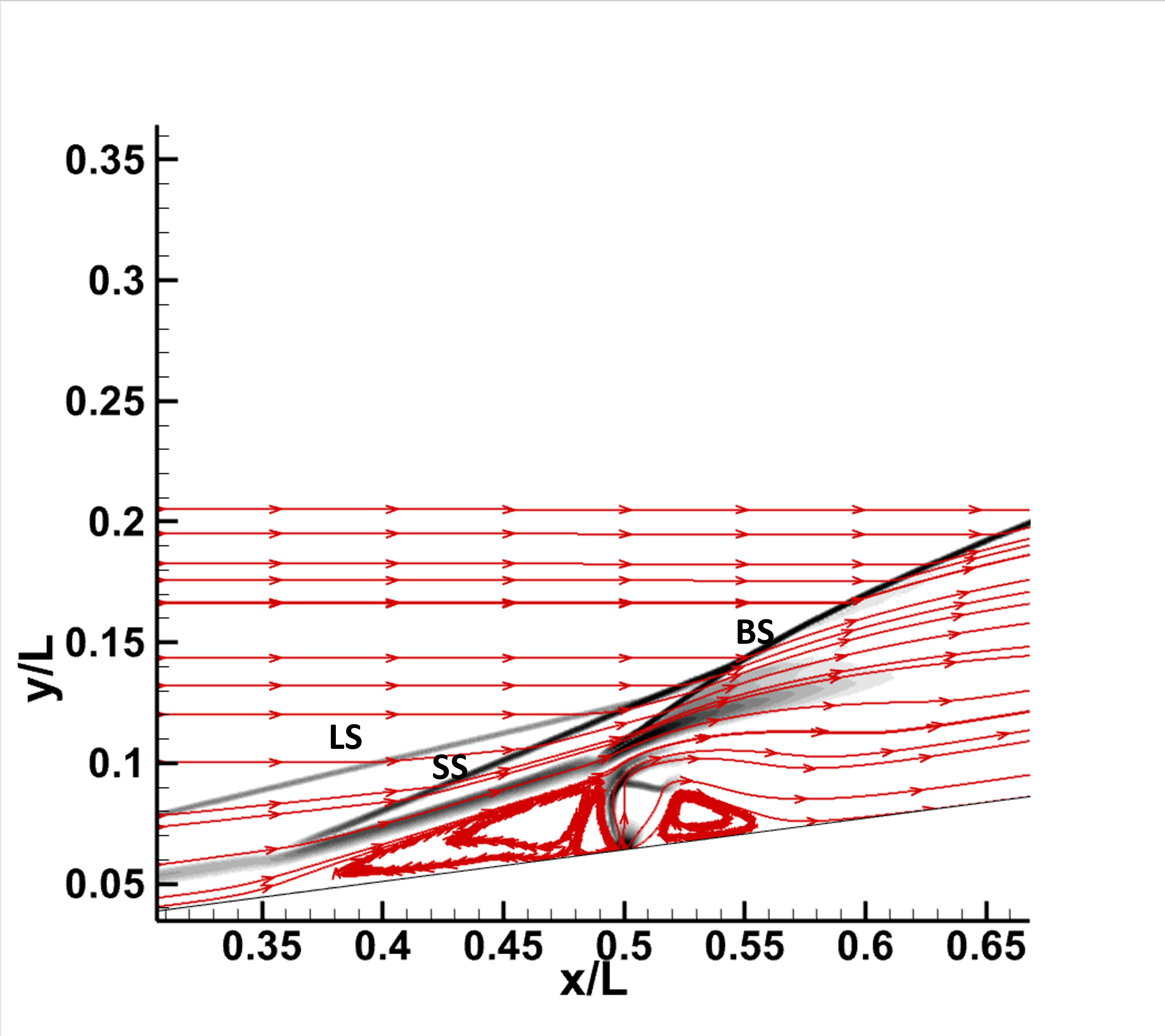}}
\subfigure[M12 case]{\label{fig:SchlierenM12}\includegraphics[trim=0cm 0cm 0cm 9.3cm,clip,width=0.48\linewidth]{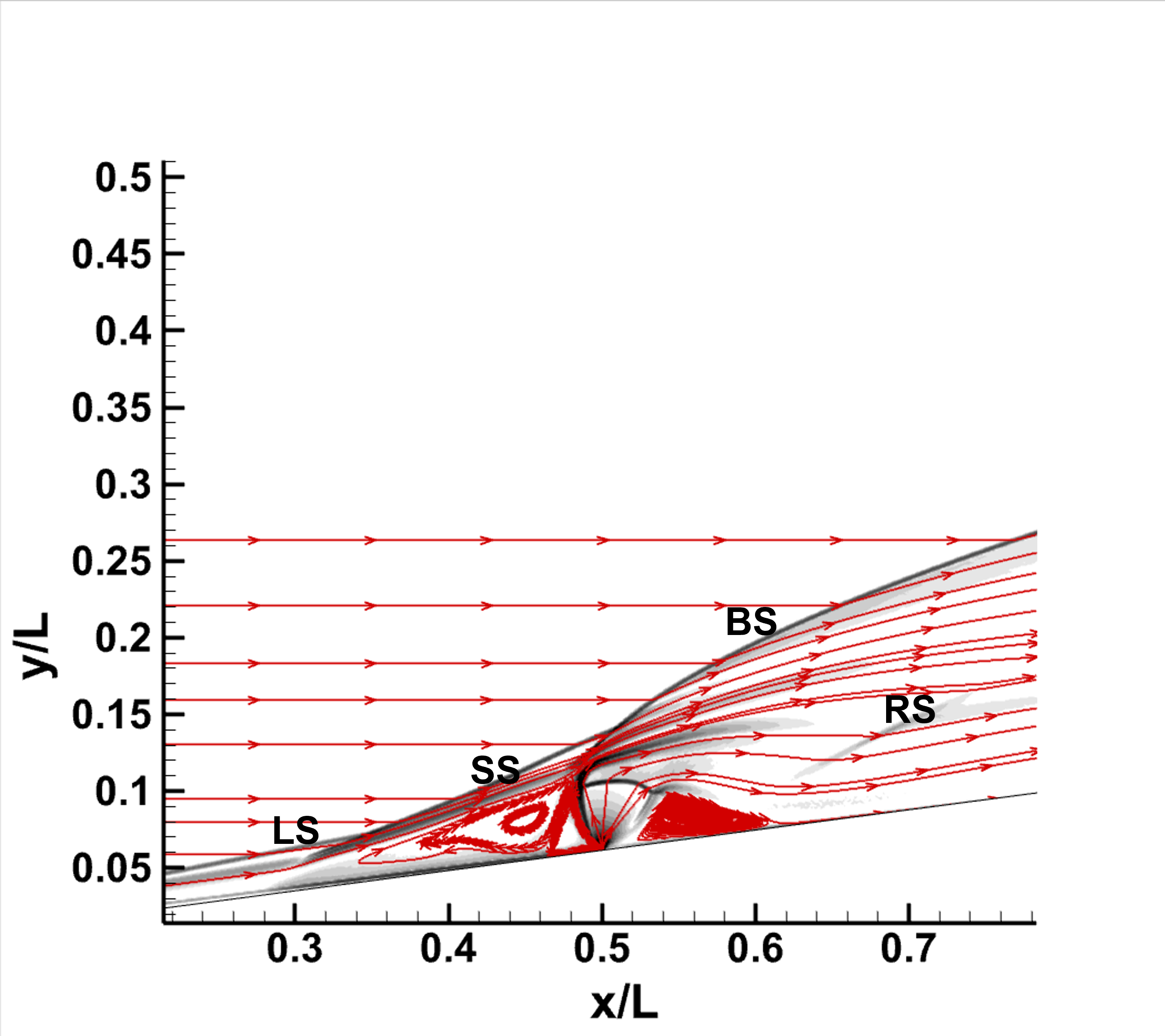}}
\end{center}
\caption{Contours of Mach number and numerical Schlieren for M6 and M12 cases.}
\label{fig:DSMCGroundStateResults}
\end{figure}

First we consider the flow features that occur in time averaged flowfields, as shown in Fig.~\ref{fig:DSMCGroundStateResults}.   It can be clearly seen from both Mach number contours with streamlines (Figs.~\ref{fig:MaContourM6} and ~\ref{fig:MaContourM12}) and numerical Schlieren images (Figs.~\ref{fig:SchlierenM6} and ~\ref{fig:SchlierenM12}) that the characteristic structures in the side jet interaction zones discussed by Cassel ~\cite{CasselReview2003} are  well captured. Starting with the Mach number contours, two main recirculation zones are identified: one upstream (UR) and one downstream (DR) of the jet. The upstream recirculation zone is caused by the separation of the incoming boundary layer upstream of the jet and also consists of two smaller recirculation zones (UR-1 and UR-2). It is also notable that these smaller recirculation regions in the upstream separation zone are unequal in size which is in agreement with the earlier works of Refs.~\cite{cassel1973,CasselReview2003}. The main features of the characteristic shock structure surrounding the interaction region is also seen, starting from the leading edge shock (LS) that is created by the tip of the cone. Due to the separation caused by the presence of the jet, a separation shock (SS) and a separated shear layer (SL) forms upstream of the jet. A bow shock (BS) caused by the jet expansion and reattachment shock (RS), where the flow reattaches downstream of the jet, are also seen. These shocks create Edney type-IV shock-shock interactions~\cite{edney1968} which are well known to be unsteady, as we showed in our earlier DSMC simulations~\cite{tumukluPoF,tumukluPoF2,sawant_theofilis_levin_2022}.

\begin{figure}[htbp]
\begin{center}
\subfigure[M=6 case]{\label{fig:CBParameterM6}\includegraphics[trim=0cm 0cm 0cm 9cm,clip,width=0.48\linewidth]{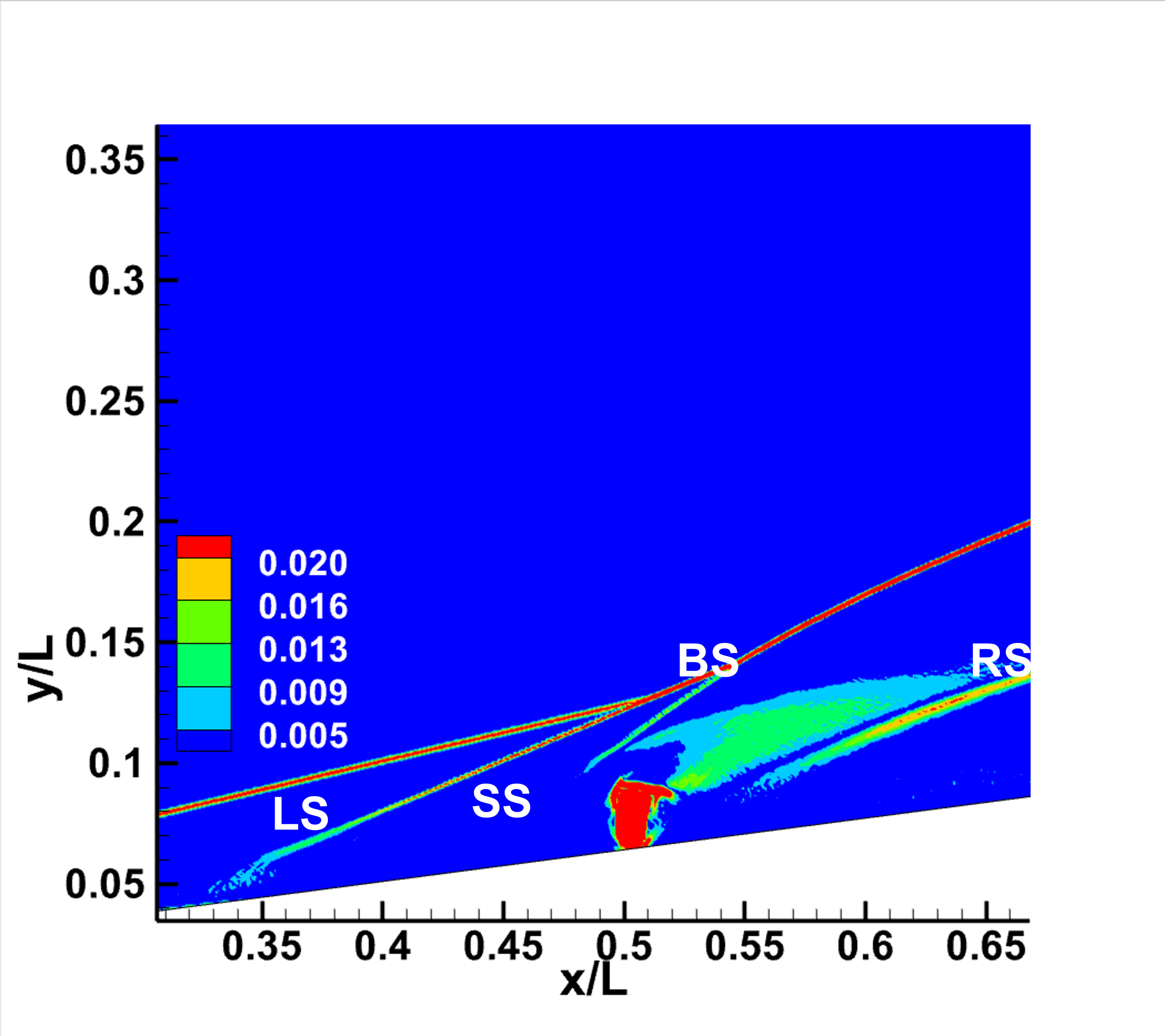}}
\subfigure[M=12 case]{\label{fig:CBParameterM12}\includegraphics[trim=0cm 0cm 0cm 9.3cm,clip,width=0.48\linewidth]{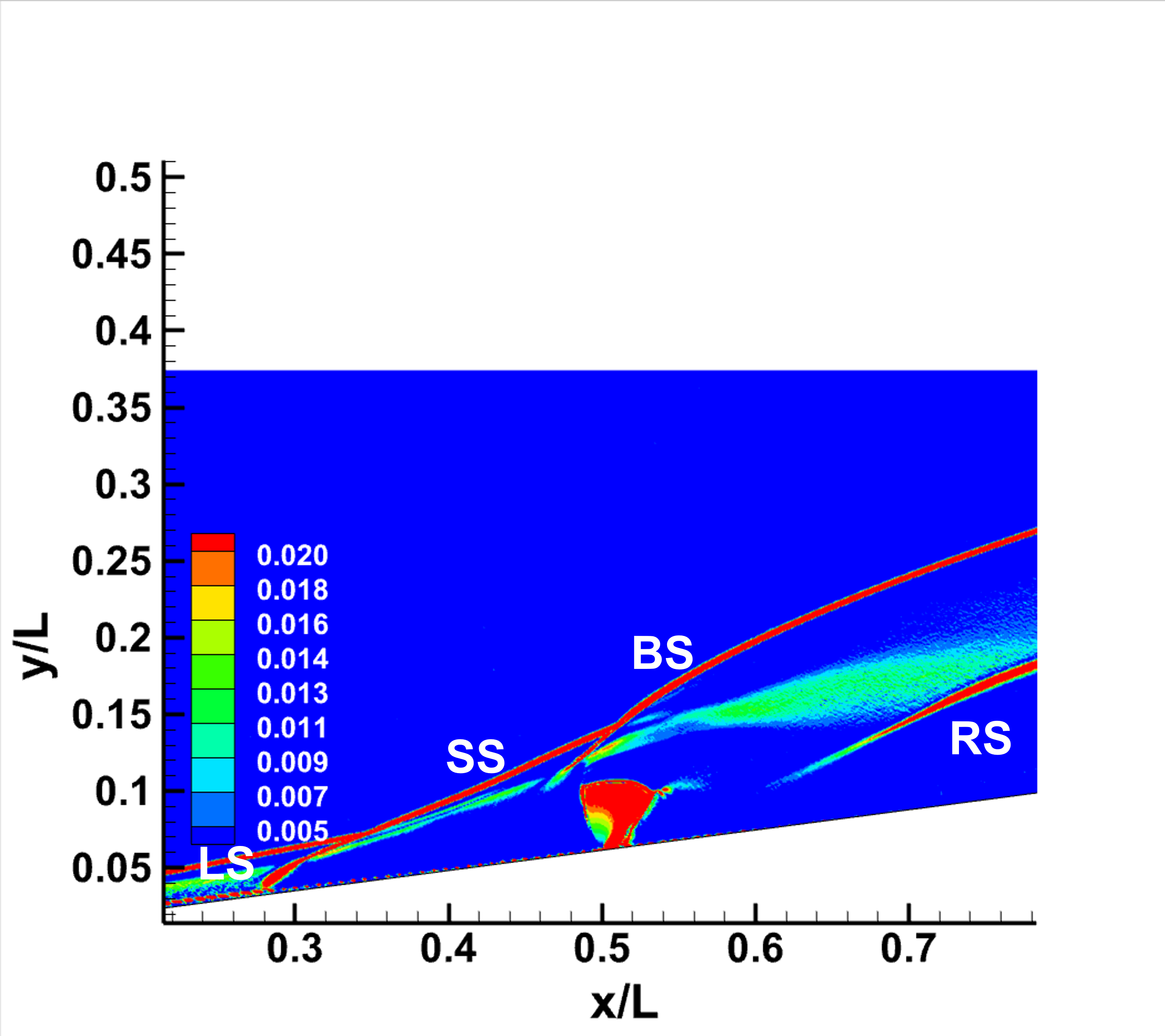}}
\end{center}
\caption{Contours of continuum breakdown parameter for M6 and M12 cases. The maximum contour level is 0.02 which indicates that continuum breakdown occurs in the red colored regions.}
\label{fig:BreakdownParameters}
\end{figure}

As mentioned earlier, an important consideration of flows dominated by shocks, expansions and shear layers is the phenomena of continuum breakdown, or regions in the flow where the continuum assumption is not valid {\it regardless} of the free stream Reynolds number.  Bird defined a continuum breakdown (CB) parameter ($P$) that depends mainly on density gradient, Ma number, and local mean free path and is given by ~\cite{Bird}
\begin{equation}
\label{eqn:breakdown}
P = \sqrt {\frac{{\pi \gamma }}{8}} \frac{{M\lambda }}{\rho }\left| {\frac{{d\rho }}{{d\overline x }}} \right|
\end{equation}
According to this CB parameter $P$, Bird estimated that continuum breakdown occurs when $P>0.02$. Since density gradient is an important term when considering continuum breakdown, not only the rarefied flows have continuum breakdown, but it is seen in any flow with a sufficiently high density gradient.  
 In Fig.~\ref{fig:BreakdownParameters} which shows the spatial distribution of the continuum breakdown parameter contours in the interaction zone, it is 
  clear that in the leading edge, separation, and bow and reattachment shocks continuum breakdown occurs. In addition, and of even more importance, continuum breakdown occurs in the jet expansion. The jet expands from the sonic outlet to a Mach number of 5-6 in a very small distance thus causing very high gradients even though the jet is at a higher number density than the incoming flow.

\begin{figure}[htbp]
\begin{center}
{\includegraphics[trim=0cm 0cm 0cm 0cm,clip,width=0.70\linewidth]{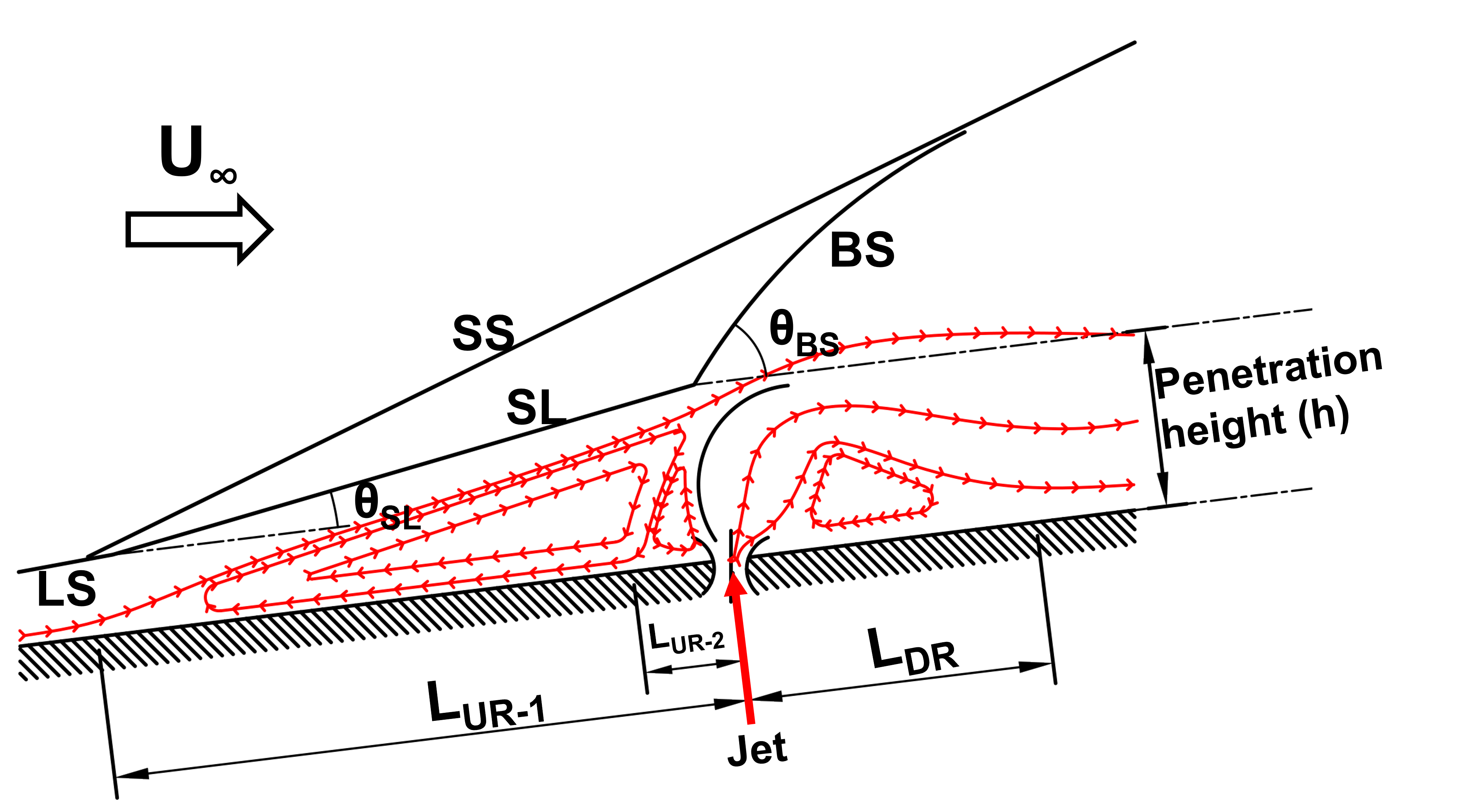}}
\end{center}
\caption{Sketch of the interaction region showing shear layer angle $\theta_{SL}$, bow shock angle $\theta_{BS}$, penetration height and upstream ($L_{UR-1}$,$L_{UR-2}$) and downstream ($L_{DR}$) separation lengths.}
\label{fig:DefineParameters}
\end{figure}

We now turn our attention to the separation and reattachment lengths and their relation to the jet penetration height. The definition of the penetration height is given in Fig.~\ref{fig:DefineParameters} following Ref.~\cite{zukowski1964}. The penetration height is defined as the normal distance from the surface to the location where there is a black "streak" between the bow shock and the normal shock of the jet expansion in the Schlieren images~\cite{zukowski1964,Rosenbaum1975}. This streak is an increase in the Schlieren intensity where the jet flow is turned by the free stream and is assumed to be the furthest that the jet gas can penetrate into the flow. The results for the M=6 and M=12 cases are presented in Table~\ref{tab:SepReattachM6M12}. It can be seen that the primary upstream separation length $L_{UR-1}$, defined as the distance along the cone surface from the jet location to the separation point as shown in Fig.~\ref{fig:DefineParameters}, is 3.5-4 times the penetration height for both cases. Furthermore, both the secondary separation $L_{UR-2}$ and downstream reattachment $L_{DR}$ lengths scale with the penetration height, as expected, and are less sensitive to the free stream parameters. 

Several other flow features such as,  dead air zone angles, bow shock~\cite{Mair1952}, penetration height~\cite{Rosenbaum1975}  and amplification factor~\cite{cassel1973}, describe the flowfield in the interaction region and allow us to relate our simulations with the existing literature.  The definition of these angles and their relationship to the penetration height are again given in Fig.~\ref{fig:DefineParameters}. The amplification factor, $K_F$, is defined as the ratio of total interaction force, {\em i.e.},  thrust from the jet plus  surface forces due to the interaction region, to the vacuum jet thrust and is given as,
\begin{equation}
\label{eqn:ampFactor}
{K_F} = \frac{{{F_{Interaction}} + {F_{Thrust}}}}{{{F_{VacuumThrust}}}}
\end{equation}
 First, we compare the non-dimensionalized penetration heights that were shown in Table~\ref{tab:SepReattachM6M12} with theoretical relationships taken from Refs.~\cite{Rosenbaum1975} and \cite{zukowski1964}. The theoretical penetration height may be expressed as,
 \begin{equation}
\label{eqn:penetheight}
{\raise0.7ex\hbox{$h$} \!\mathord{\left/
 {\vphantom {h {{r_b}}}}\right.\kern-\nulldelimiterspace}
\!\lower0.7ex\hbox{${{r_b}}$}} = 2{\left[ {\frac{{{\gamma _j}\widetilde m}}{{{U_\infty }}}} \right]^{{\raise0.7ex\hbox{$1$} \!\mathord{\left/
 {\vphantom {1 2}}\right.\kern-\nulldelimiterspace}
\!\lower0.7ex\hbox{$2$}}}}{\left[ {\frac{{2{R_j}{T_{0j}}}}{{{\gamma _j}({\gamma _j} - 1)}}} \right]^{{\raise0.7ex\hbox{$1$} \!\mathord{\left/
 {\vphantom {1 4}}\right.\kern-\nulldelimiterspace}
\!\lower0.7ex\hbox{$4$}}}}
\end{equation}
where the scaled mass flow rate, $\widetilde m $, is given by normalizing the jet mass flow rate by the free stream velocity, density and base area of the cone, {\em i.e.,}
\begin{equation}
\label{eqn:normalizedmass}
\widetilde m = (\dot m/4{\rho _\infty }{U_\infty }{A_b})
\end{equation}
Then using the scaled mass flow rate,  the theoretical penetration height given by Eq.~\ref{eqn:penetheight}, may be obtained, where the equation is  derived from the momentum balance of the jet expansion.  Note that to calculate the mass flow rate from the jet, we have used its circumferential outlet area since it is an axisymmetric case.  It can be seen from the first row of Table~\ref{tab:DSMCvsTheory} that our results agree very well with the theoretical calculations for the penetration height. 
Next  we  consider  angles of the shear layer and bow shock with the cone surface, since they significantly influence shock-shear layer interactions. Using Fig.~10 of Ref.~\cite{Mair1952} we compare the bow shock angles. Note that for each shear layer angle a bow shock angle can be calculated from Fig.~10 of Ref.~\cite{Mair1952} or vice versa. We take the shear layer angles from our simulations to calculate the expected bow shock angles. Our results again agree very well with the expected shock angles, as shown in Table ~\ref{tab:DSMCvsTheory}.  Lastly, we compare the amplification factor with those given in Ref.~\cite{cassel1973}. In terms of amplification factors most experimental results found in the literature range from supersonic to hypersonic and from laminar to turbulent flows. The most relevant cases for this work are the hypersonic and laminar cases with amplification factors between 2.5-5.2. In our two cases we obtain amplification factors that are higher than this range, mainly due to our use of a jet mass flow rate smaller than the cases in the literature. However, even for the smaller mass flow rate, the jet causes enough blockage to create a large interaction region because our cases are axisymmetric. This has the effect of causing the the interaction force to be much higher than that from the jet thrust.

\begin{table}[htb!]
\caption{Penetration height, and separation and reattachment distances for the M=6 and M=12 cases. Separation and reattachment distances are non-dimensionalized by the penetration height for each case.}
\label{tab:SepReattachM6M12}
\begin{center}
\scalebox{1.0}{
\begin{tabular}{cccccccccc}
\hline
\hline
Parameter&M6&M12\\
\hline
Penetration height, h (m) &0.015&0.025\\
Primary upstream separation length ($L_{UR-1}/h$) &4.02&3.32\\
Secondary upstream separation length ($L_{UR-2}/h$) &0.67&0.55\\
Downstream reattachment length ($L_{DR}/h$) &1.67&1.81\\
\hline
\hline
%\end{tabularx}
\end{tabular}}
\end{center}
\end{table}

\begin{table}[htb!]
\caption{Comparison of the parameters from DSMC with the literature.}
\label{tab:DSMCvsTheory}
\begin{center}
\scalebox{0.8}{
\begin{tabular}{cccccccccc}
\hline
\hline
 &M6&&M12& \\
\hline
Parameter &DSMC&Literature~\cite{Mair1952,cassel1973,Rosenbaum1975}&DSMC&Literature~\cite{Mair1952,cassel1973,Rosenbaum1975} \\
\hline
Penetration height (h/r$_{b}$)~\cite{Rosenbaum1975} &0.32 &0.331 &0.5 &0.523\\
Shear layer ($\theta_{SL}$), bow shock angles ($\theta_{BS}$)~\cite{Mair1952} &17, 33 &17, 36&13, 31 &13, 33 \\
Amplification factor ~\cite{cassel1973}  &5.9 &2.5-5.2 &2.89 &2.5-5.2  \\
\hline
\hline
%\end{tabularx}
\end{tabular}}
\end{center}
\end{table}

\begin{figure}[htb!]
\begin{center}
\subfigure[Mole fraction and streamlines of  jet gas, M=6 case]{\label{fig:fracM6}\includegraphics[clip=true,trim=02 02 02 510,width=0.48\textwidth]{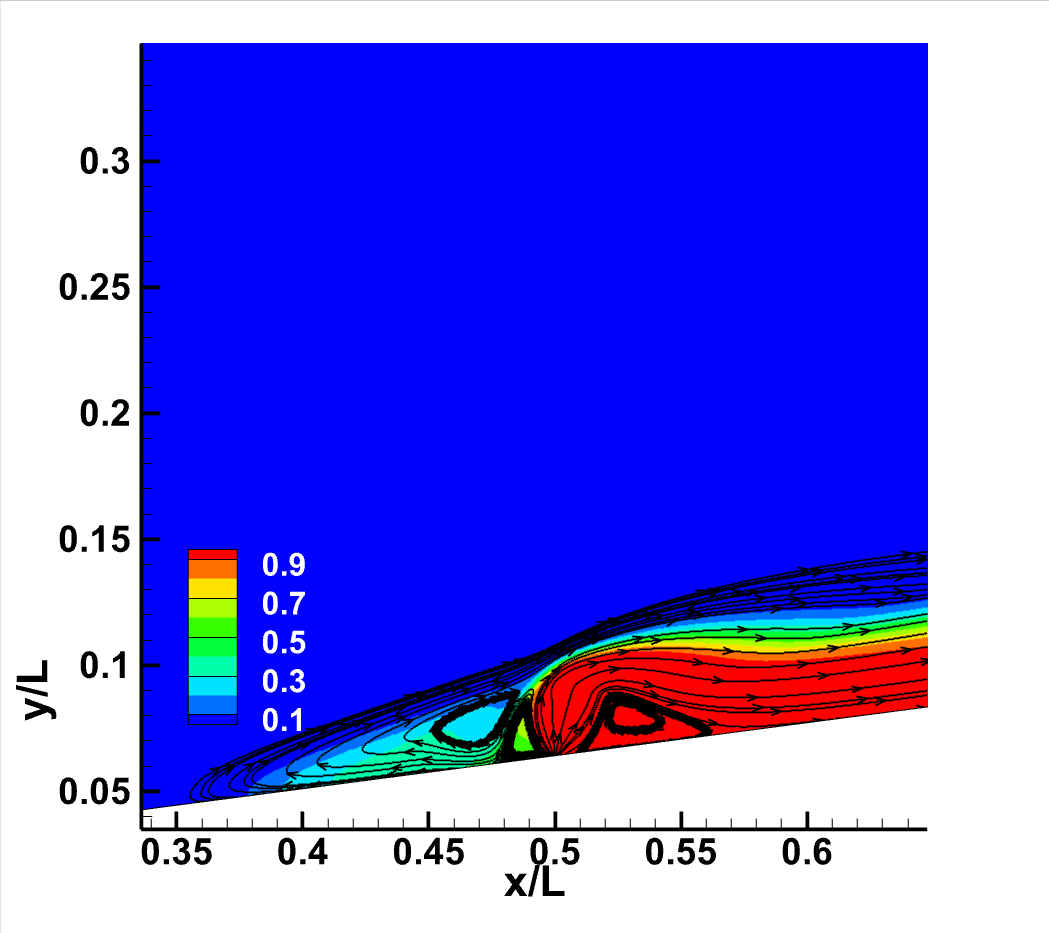}}
\subfigure[Mole fraction and streamlines of  jet gas, M=12 case]{\label{fig:fracM12}\includegraphics[clip=true,trim=02 02 02 510,width=0.48\textwidth]{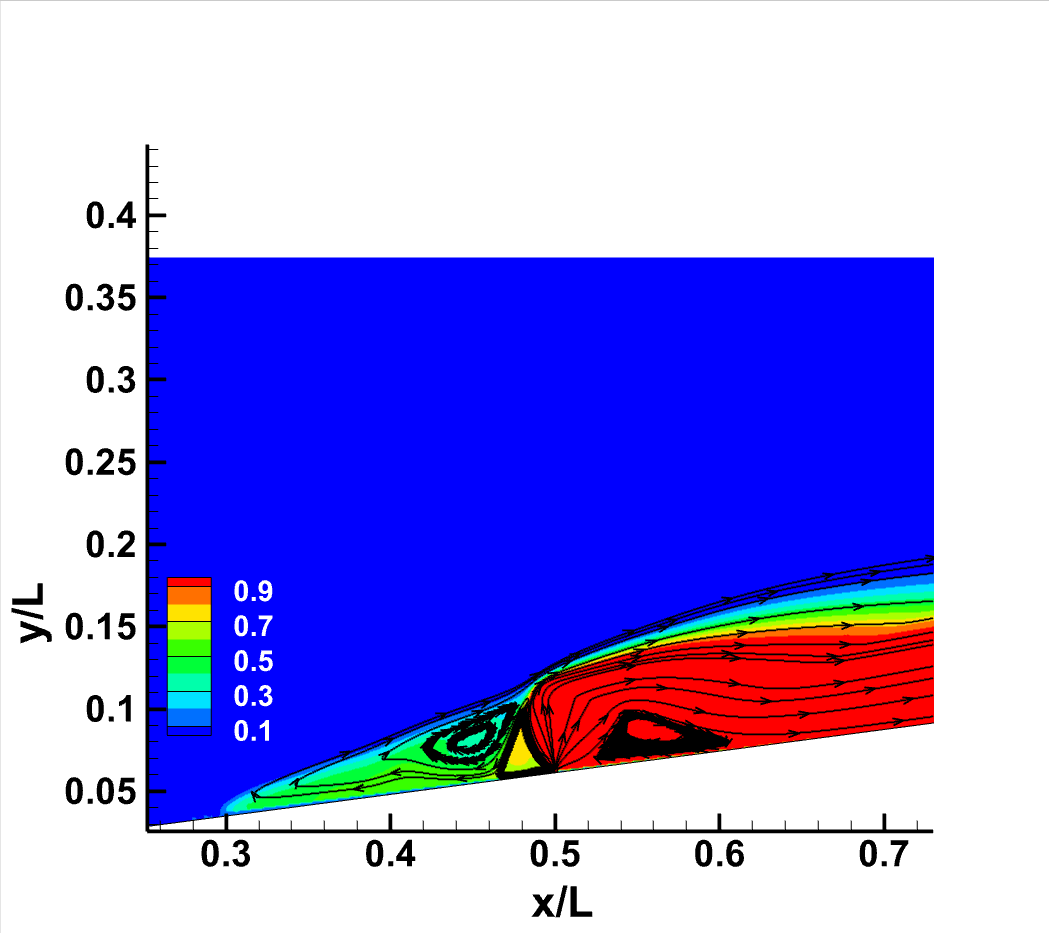}}
\subfigure[Translational temperature, M=6 case]{\label{fig:TtrnM6}\includegraphics[clip=true,trim=02 02 02 250,clip,width=0.48\linewidth]{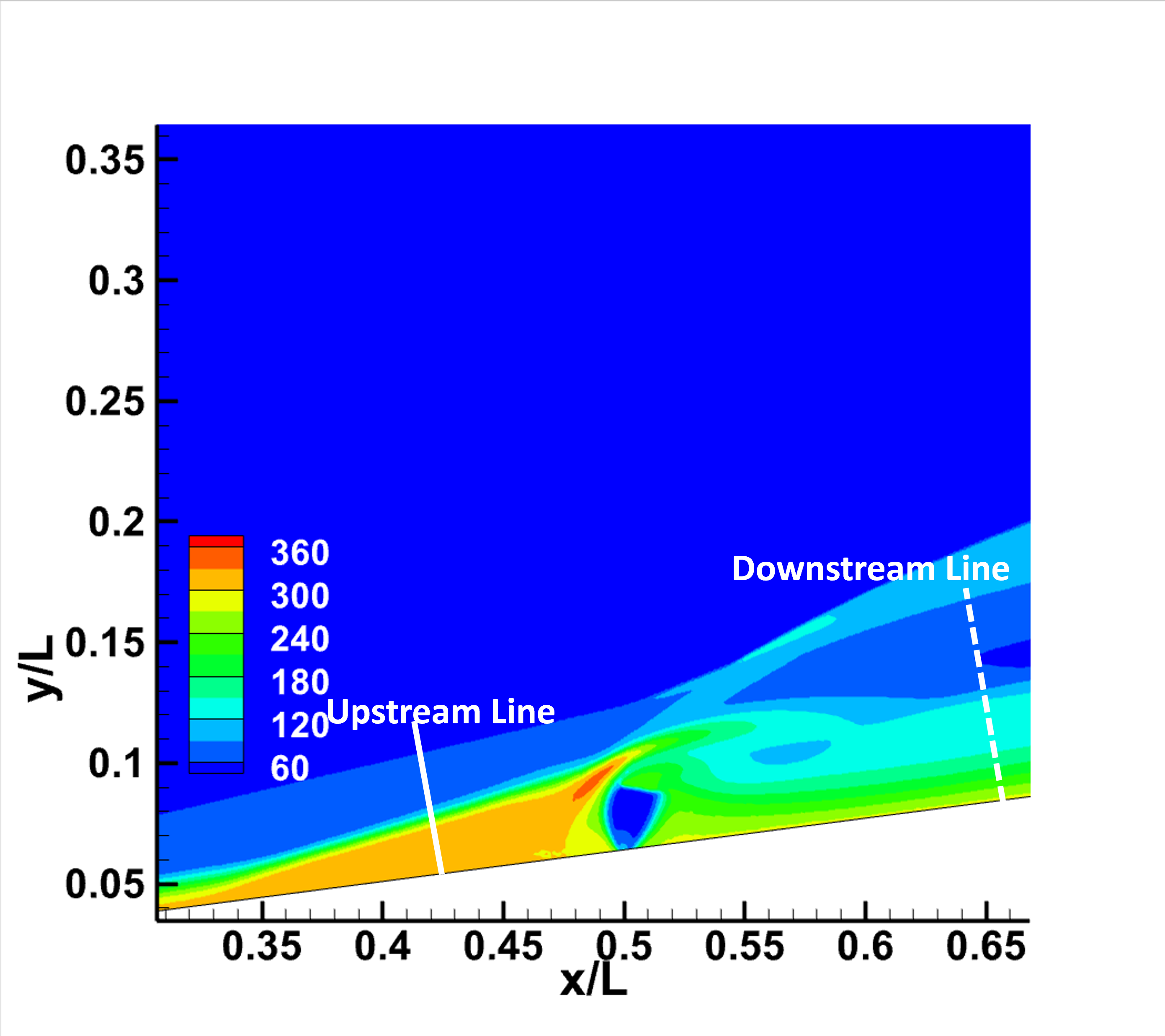}}
\subfigure[Translational temperature, M=12 case]{\label{fig:TtrnM12}\includegraphics[clip=true,trim=02 02 02 240,clip,width=0.48\linewidth]{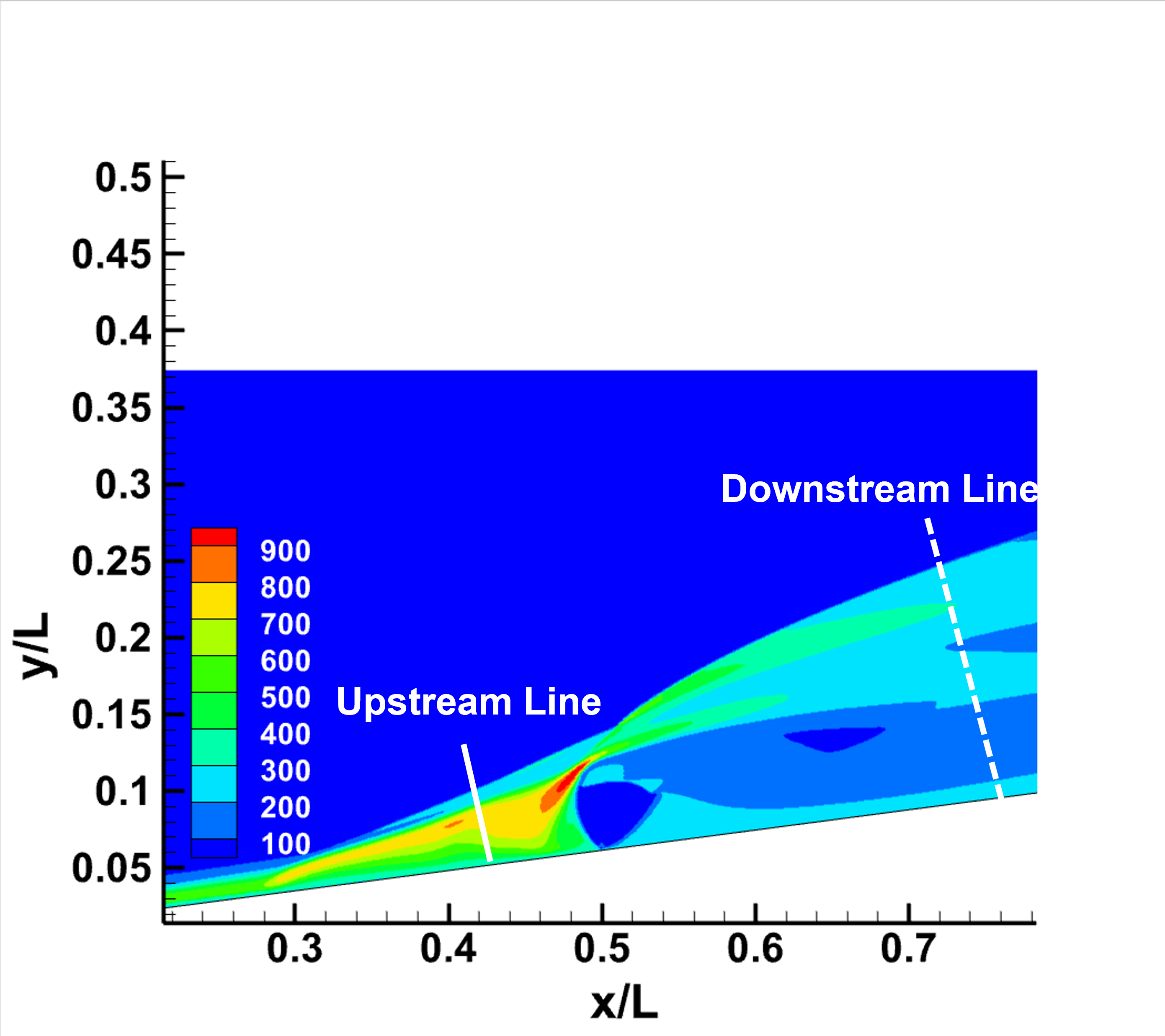}}
\subfigure[Translational temperature and mole fraction profiles, M=6 case.]{\label{fig:MoleTtrnM6BL}\includegraphics[clip=true,trim=0.5cm 0.5cm 0.5cm 0.5cm,clip,width=0.48\linewidth]{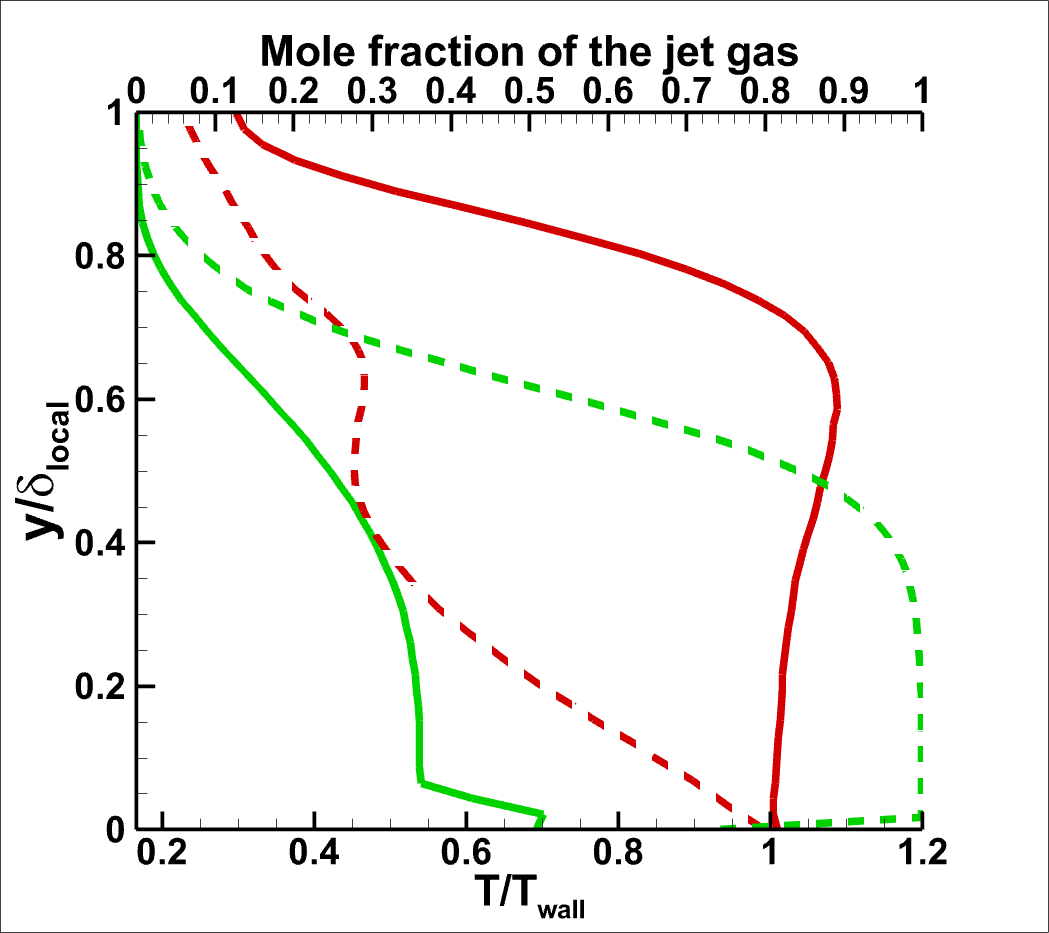}}
\subfigure[Translational temperature and mole fraction profiles, M=12 case.]{\label{fig:MoleTtrnM12BL}\includegraphics[clip=true,trim=0.5cm 0.5cm 0.5cm 0.5cm,clip,width=0.48\linewidth]{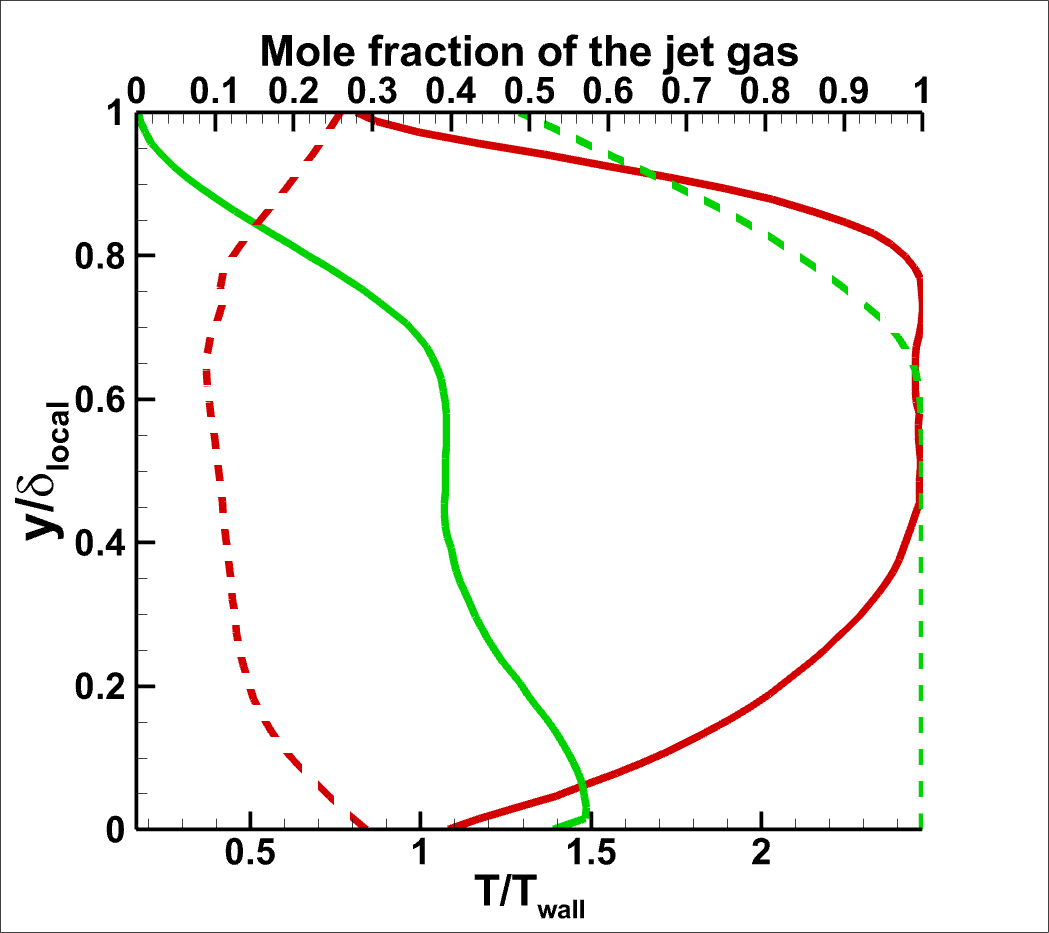}}
\end{center}
\caption{Mole fraction of jet gas and translational temperature contours. Also shown in the last row: translational temperature and mole fraction profiles along the white lines shown in middle row figures at locations upstream and downstream of the jet. Solid lines: upstream location, dashed lines: downstream location, red lines: temperatures and green lines: mole fractions.}
\label{fig:TTrnFracMixing}
\end{figure}

The mixing characteristics of the flow are another important effect created by the presence of the side jet. Figs.~\ref{fig:fracM6} and ~\ref{fig:fracM12} show the mole fractions of jet gas for the two Mach number cases where it can be seen that the only mixing zones are the upstream separation zones UR-1 and UR-2 in both cases.  The mole fraction contours demonstrate that 
 there is no mixing downstream of the jet and that the boundary layer only consists of the jet gas. To further illustrate the lack of  mixing between the jet and free stream gas, translational temperature contours are shown in Figs.~\ref{fig:TtrnM6} and ~\ref{fig:TtrnM12} where it can be seen that the jet expands to very low temperatures for both flow Mach number cases. The temperatures inside the upstream separation region are much lower than the free stream total temperatures ($T_0=420K$ for M6 case and $T_0=1500K$ for the M12 case) because of the mixing 
  of the cold jet gas with the incoming hot free stream gas.  The translational temperature of the jet gas is lower than the isothermal wall temperature of 300~K in the boundary layer downstream of the jet since there is no mixing and only the jet gas exists in the boundary layer. In Figs.~\ref{fig:MoleTtrnM6BL} and ~\ref{fig:MoleTtrnM12BL} it can be seen that temperatures decrease towards the cone surface for the upstream locations, but increase towards the wall for the downstream locations for both M6 and M12 cases. This means that to satisfy the isothermal wall boundary condition downstream of the jet, the flow must remove energy from the wall. This will have implications for the stability of the boundary layer downstream of the jet since there is energy addition to the flow which might eventually cause transition to turbulence, but that is beyond the scope of this work. 

\section{Steady State Gas-Surface Interaction Results}\label{sec:timeAveSurface}
 
An important objective of a side jet is to create forces on control surfaces. However, in a hypersonic incoming flow,  complex flow interactions due to the presence of the jet can create surface forces and influence the local heat transfer values in addition to the force from the jet itself.   Therefore for design purposes,   surface parameters such as pressure distributions, heat fluxes and friction forces in addition to the thrust force created by the jet, are useful.   

\begin{figure}[htbp]
\begin{center}
{\includegraphics[trim=0cm 0cm 0cm 0cm,clip,width=0.70\linewidth]{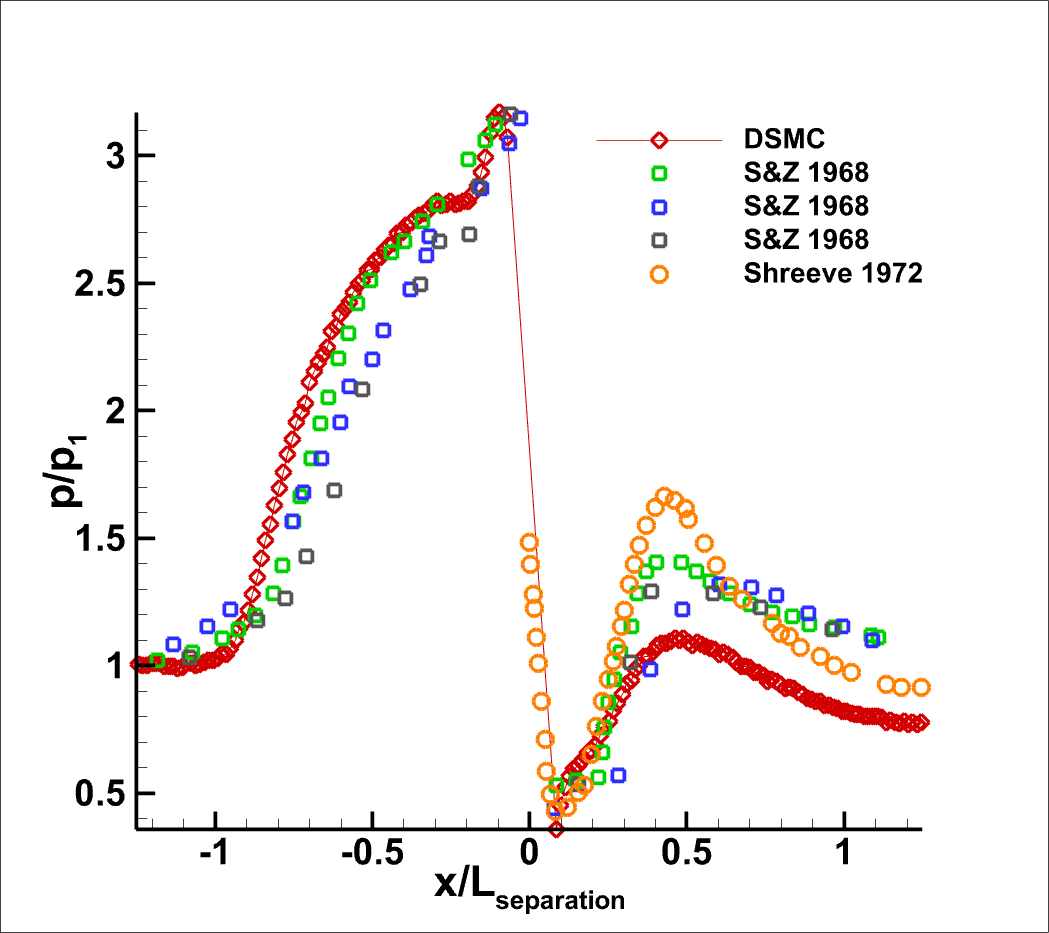}}
\end{center}
\caption{Comparison of surface pressures for M=6 case with the experimental results available in literature\cite{spaid1968,Shreeve1972}. Diamond symbols from the DSMC simulations of this work, square symbols from Spaid\&Zukoski\cite{spaid1968} denoted as "S\&Z 1968" and circle symbols are from Shreeve et. al\cite{Shreeve1972}.}
\label{fig:PressDistComp}
\end{figure}

 We first compare our simulations at M=6 with  surface pressure distributions from the experiments of Spaid\&Zukoski\cite{spaid1968} and Shreeve et.al.\cite{Shreeve1972} for 2D and axisymmetric configurations, respectively. In these experiments, the jet is either a slit along the transverse direction or a slit along the surface of the cone, thus creating nominally 2D or axisymmetric interactions. Since our simulations are also for an axisymmetric setting, these provide the best comparison in terms of the flow topology. The incoming Mach number immediately upstream of the jet in Spaid\&Zukoski\cite{spaid1968} is 4.54, comparable to our case of 5.1. Three set of pressure distributions shown in Fig.\ref{fig:PressDistComp} for different jet strengths that are between 60-200 are compared with our case which has a value of 100. The jet strength is defined in both of these experiments as the ratio of the jet total pressure to the static pressure of the incoming flow just upstream of  separation. The experiments of Shreeve et.al.\cite{Shreeve1972} are for a 5 half angle cone at M=6 free stream flow (the same Mach number as our case), and for a jet strength of 42. In Fig.~\ref{fig:PressDistComp}, we normalize the x-axis by the separation length, which is determined as the first point that the  surface pressure start to rise in  the experimental data and surface pressures are normalized by the static pressure just upstream of the separation for each case, as was done in these papers. Since Shreeve et.al.\cite{Shreeve1972} paper only gives the pressure distribution downstream, we used downstream separation length for the scaling of the data from this paper. This point is determined by the point where the pressure peaks downstream of the jet. It should be also noted that jet is now offset to origin for this figure only to have a better comparison with the experiments. It is clear from Fig.~\ref{fig:PressDistComp} that the pressure distributions obtained from the DSMC simulations  agree quite well with experimental results. In addition to the distributions, the plateau and peak pressures for the upstream and smallest pressure downstream of the jet also agree well. 

\begin{figure}[htb!]
\begin{center}
\subfigure[Surface pressure distribution]{\label{fig:SurfPress}\includegraphics[trim=0cm 0cm 0cm 0cm,clip,width=0.48\linewidth]{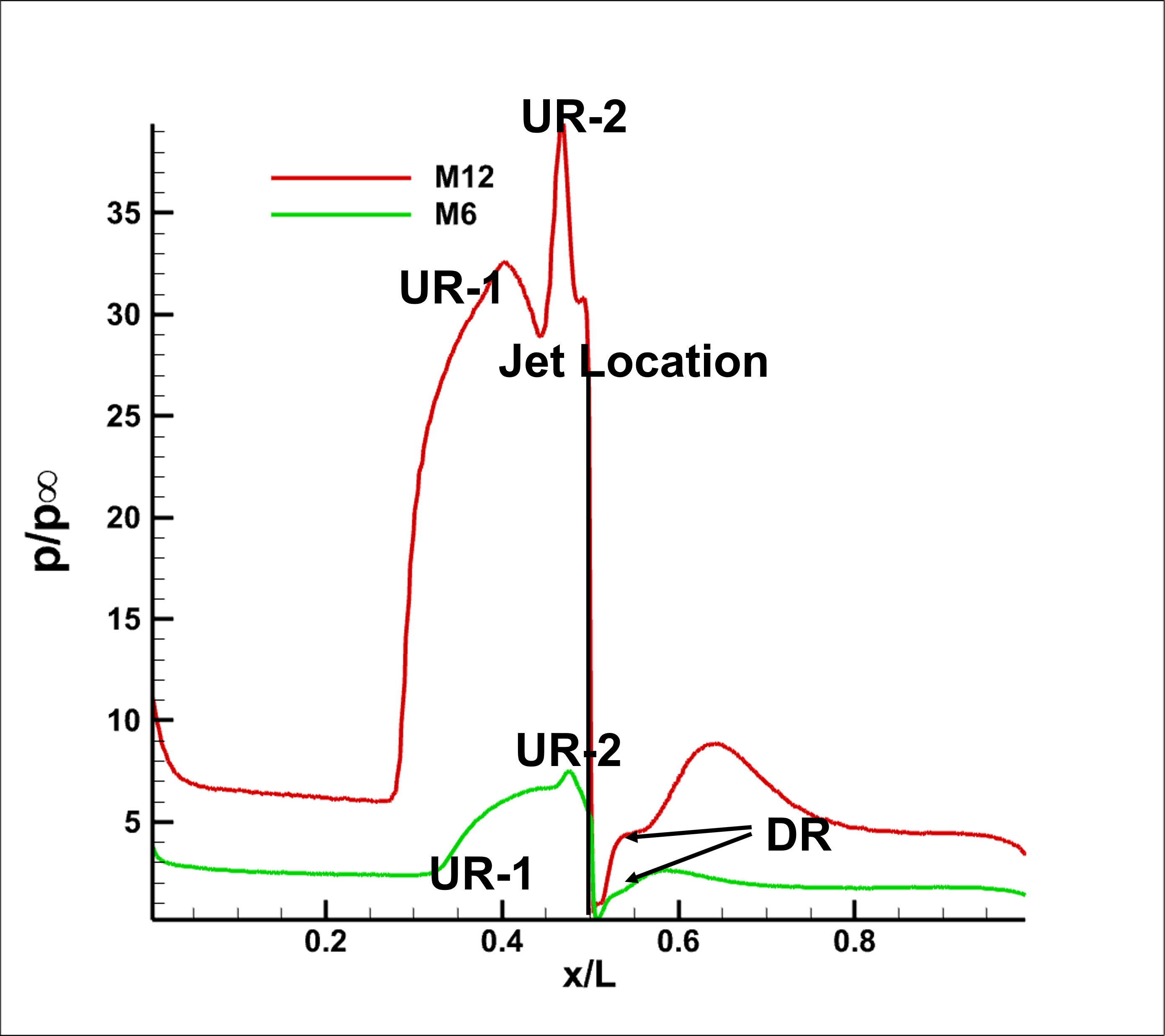}}
\subfigure[Heat transfer coefficient]{\label{fig:heatcoeff}\includegraphics[trim=0cm 0cm 0cm 0cm,clip,width=0.48\linewidth]{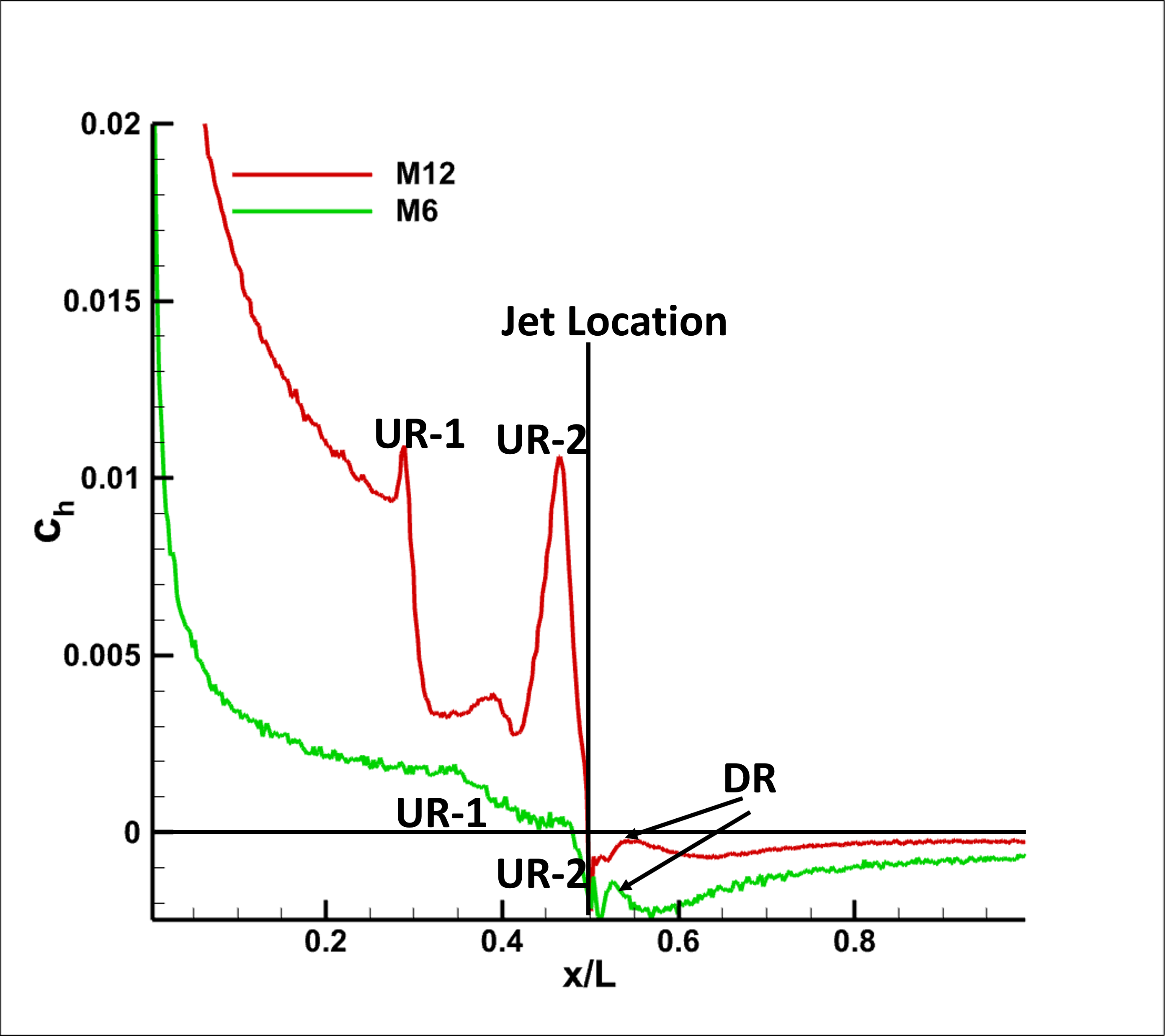}}
\subfigure[Skin friction coefficient]{\label{fig:skinfriction}\includegraphics[trim=0cm 0cm 0cm 0cm,clip,width=0.48\linewidth]{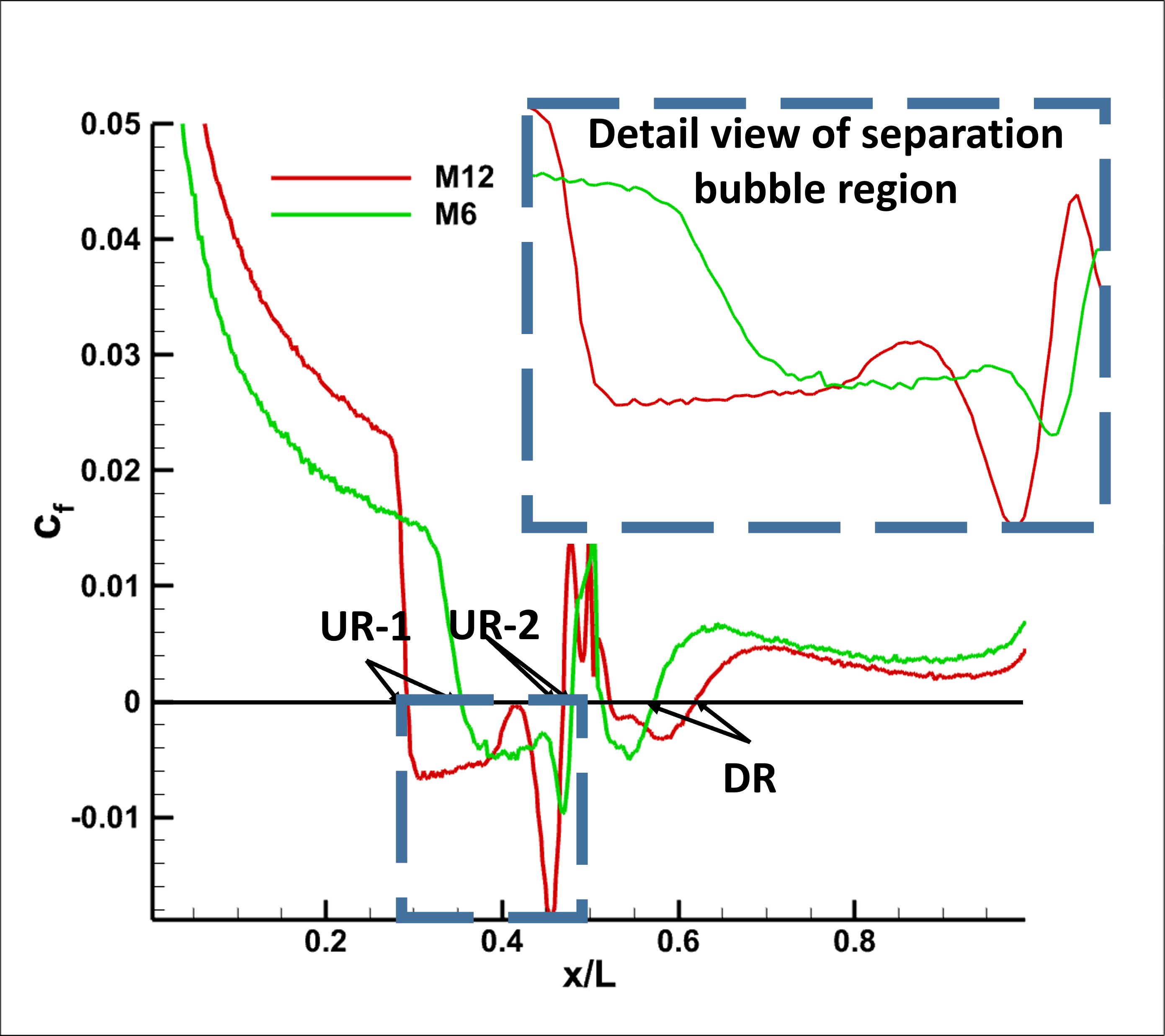}}
\caption{Time averaged pressure, heat transfer, and skin friction distributions for M=6 and M=12 cases.}
\label{fig:SurfaceDistributions}
\end{center}
\end{figure}

Having shown reasonable agreement with the available experimental data in linterature, we now show additional surface parameters for both M=6 and M=12 cases. It should be noted that in Fig.~\ref{fig:SurfaceDistributions} the x axis scaling is again with the length of the cone and the origin of the coordinate system is at the tip of the cone so that the jet is at x/L=0.5, as it was in all the previous figures except Fig.\ref{fig:PressDistComp}. It can be clearly seen from the surface parameters in Fig.~\ref{fig:SurfaceDistributions}, that due to the stronger flow interaction region for the M12 case, the maxima of all surface parameters are much higher than the M6 case. The surface pressure plateaus at $x/L=0.27$ and $x/L=0.61$ are due to the upstream and downstream separation regions which are at much higher pressure values for the M12 case compared to M6. Also, in Fig.~\ref{fig:heatcoeff} there are two heating peaks in the upstream separation region for the M12 case which are not present in the M6 case. This is mainly due to the fact that the total temperature for the M12 case is higher and the interaction of the separation shock with the leading edge shock and bow shock creates higher temperatures than the wall temperature.  More interestingly, the heat transfer rate downstream of the jet location is negative in both cases, meaning that the heat transfer is from the surface to the flow. This is due to the expansion of the jet and its mixing characteristics which were shown in Fig.~\ref{fig:TTrnFracMixing} where the boundary layer downstream of the jet only consists of the "cold" jet gas. 
 Lastly, the skin friction coefficient values  are the most similar parameter 
for the two cases, but as expected, the locations where the coefficient changes sign are different. The shear stress distributions inside the primary upstream separation zone (UR-1) are consistent with those observed previously in the large separation zones in 
 compression corners for supersonic to hypersonic Mach conditions where two minima were found to be present.
~\cite{smith_1998,korolev_etal_2002,gai_khraibut_2019}. 

The integral quantities of the above surface parameters,  total drag force, normal force, and heat transfer, were  obtained  by integrating local skin friction, pressure distributions, and 
heat transfer rate over the surface of the cone. 
%Drag force is the force acting in the opposite direction of the incoming free stream flow and the normal force is the force acting in the normal direction to the free stream flow. 
Comparisons are made for the integral force quantities for the cases with and without the jet present, and by excluding of the jet thrust force from the jet itself to understand the effect of  aerodynamic forces only. Table \ref{tab:SurfaceForces} shows the ratio of the forces with and without the jet for both drag and normal forces. It is clear that the presence of the jet interaction region decreases the overall drag on the cone. Moreover, due to the high pressure zones created by the separation regions, the normal force increases to almost twice the value of the case with no jet present for both the M6 and M12 cases.  More interestingly, we see that total heat transfer rate decreases and for the M6 case it is actually a negative value, as shown in Table \ref{tab:SurfaceHeat}. This means that the heat transfer is from the wall to the flow, and even for the M12 case the jet causes the overall heat transfer to drop by 80\% of the case with no jet present. One advantage of using DSMC is that we can track how much of these surface properties are due to each gas. In Table \ref{tab:SurfaceHeat} we show that the heat transfer due to the jet gas is much smaller (for M6 case it is even negative) compared to the free stream gas. 
 
\begin{table}[htbp]
\caption{The ratio of surface forces with and without a side jet present. The quantities given here are the integral quantities over the cone surface.}
\label{tab:SurfaceForces}
\begin{center}
\scalebox{0.8}{
\begin{tabular}{cccccc}
\hline
\hline
Force & Formula &M=6&M=12\\
\hline
Drag Force Ratio&$F_{Djet}/F_{Dnojet}$&0.76&0.68 \\
Normal Force Ratio&$F_{Njet}/F_{Nnojet}$&2.30 &1.74\\
\hline
\hline
%\end{tabularx}
\end{tabular}}
\end{center}
\end{table}

\begin{table}[htbp]
\caption{Heat transfer rates with and without a side jet present. The quantities given here are the integral quantities over the cone surface.}
\label{tab:SurfaceHeat}
\begin{center}
\scalebox{0.8}{
\begin{tabular}{llllccc}
\hline
\hline
Jet Presence&Property &M=6&M=12\\
\hline
Without jet &Total heat transfer rate(W) &12.6&249.2 \\
With jet &Total heat transfer rate(W) &-5.68&59.3 \\
With jet &Heat transfer rate due to jet gas(W)&-8.8&4.9\\
With jet &Heat transfer rate due to free stream gas(W)&3.12&54.4 \\
\hline
\hline
%\end{tabularx}
\end{tabular}}
\end{center}
\end{table}

\section{Instantaneous Flowfield Results and Unsteadiness}\label{sec:InstaFlow}

Up to this point we have discussed the effect of the jet on time averaged, steady state flowfields. As mentioned earlier, however, we anticipate that there will be a significant amount of unsteadiness present in the strong shock system, shear layers, and mixing zones.  Therefore we investigate the instantaneous flowfields and time resolved probe data taken from strategic flow locations.

\begin{figure}[htbp]
\begin{center}
\subfigure[ $t_1+\Delta t_{M6}$ ]{\label{fig:Pressdif1M6}\includegraphics[trim=0.25cm 0.5cm 0.5cm 9cm,clip,width=0.48\linewidth]{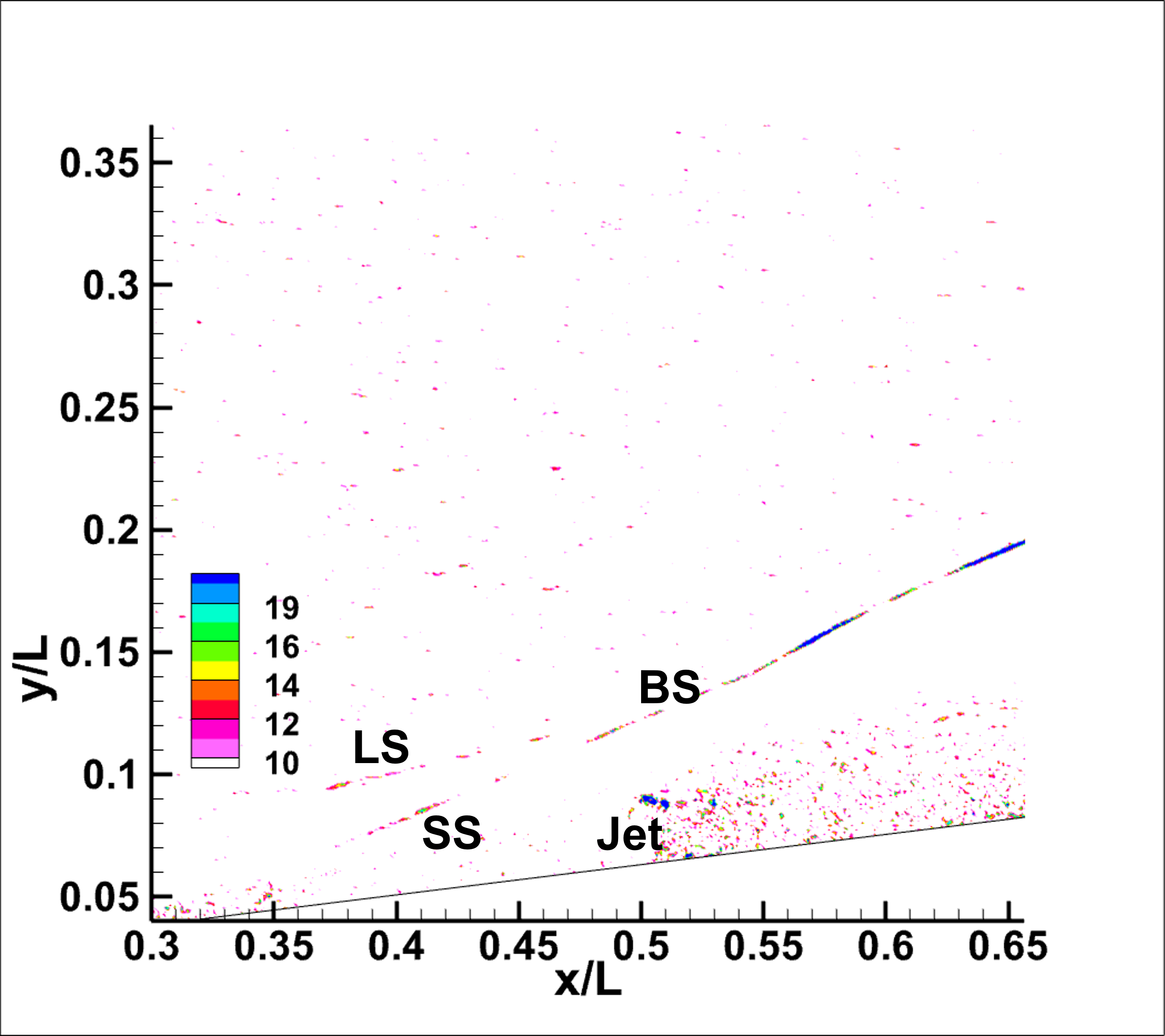}}
\subfigure[ $t_1+\Delta t_{M12}$ ]{\label{fig:Pressdif1M12}\includegraphics[trim=0.25cm 0.5cm 0.5cm 9cm,clip,width=0.48\linewidth]{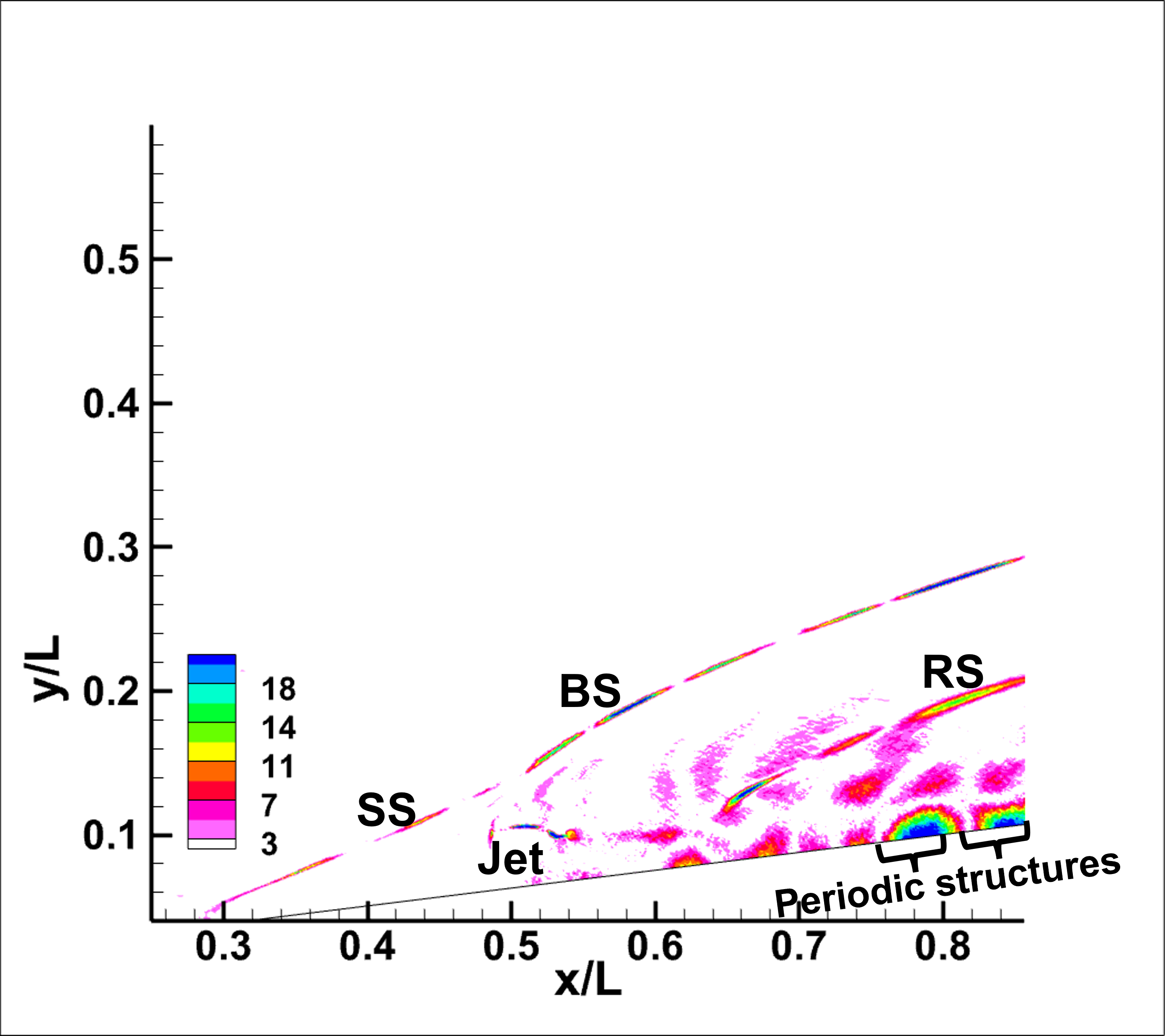}}
\subfigure[ $t_1+2\Delta t_{M6}$]{\label{fig:Pressdif2M6}\includegraphics[trim=0.25cm 0.5cm 0.5cm 9cm,clip,width=0.48\linewidth]{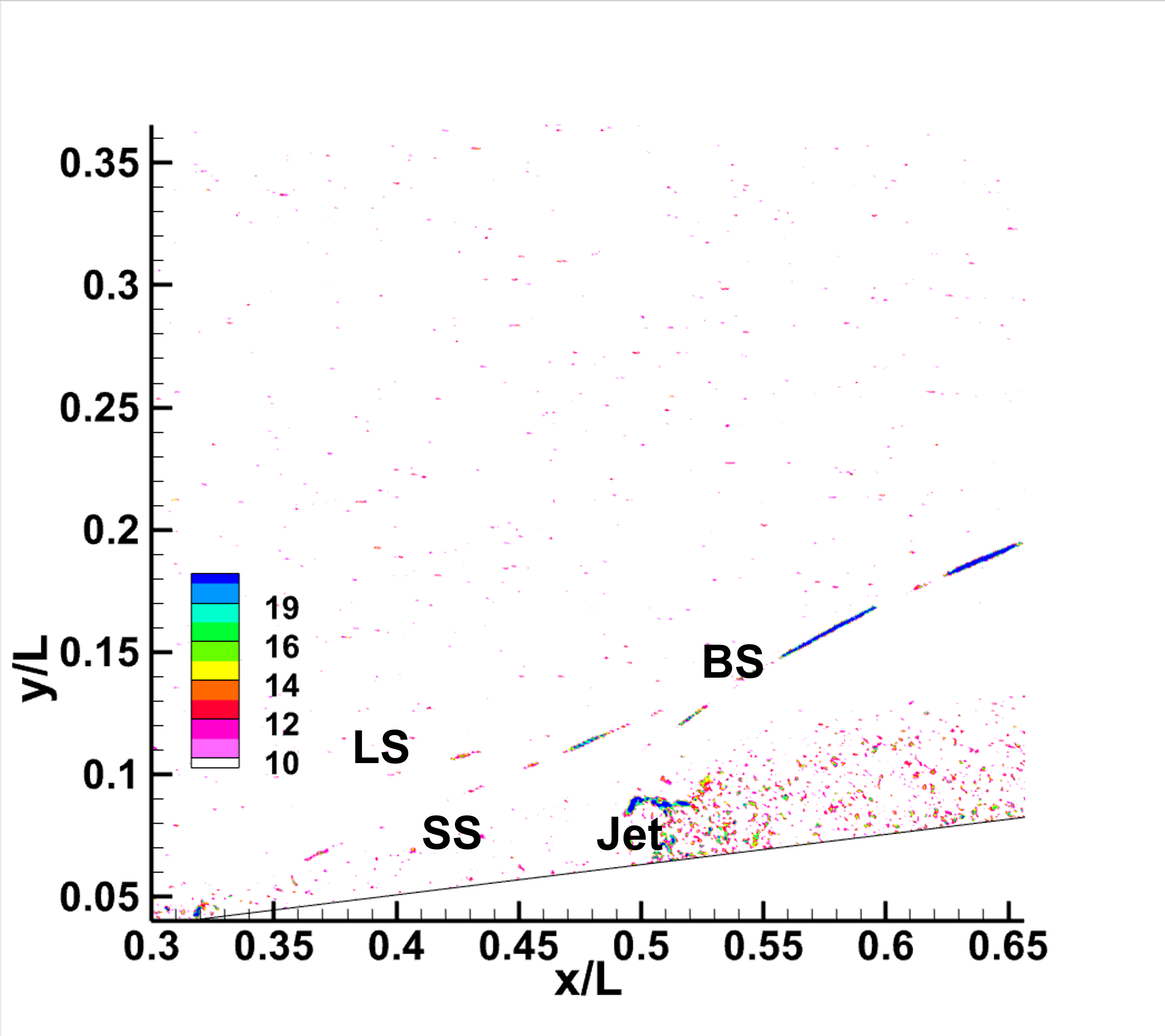}}
\subfigure[ $t_1+2\Delta t_{M12}$ ]{\label{fig:Pressdif2M12}\includegraphics[trim=0.25cm 0.5cm 0.5cm 9cm,clip,width=0.48\linewidth]{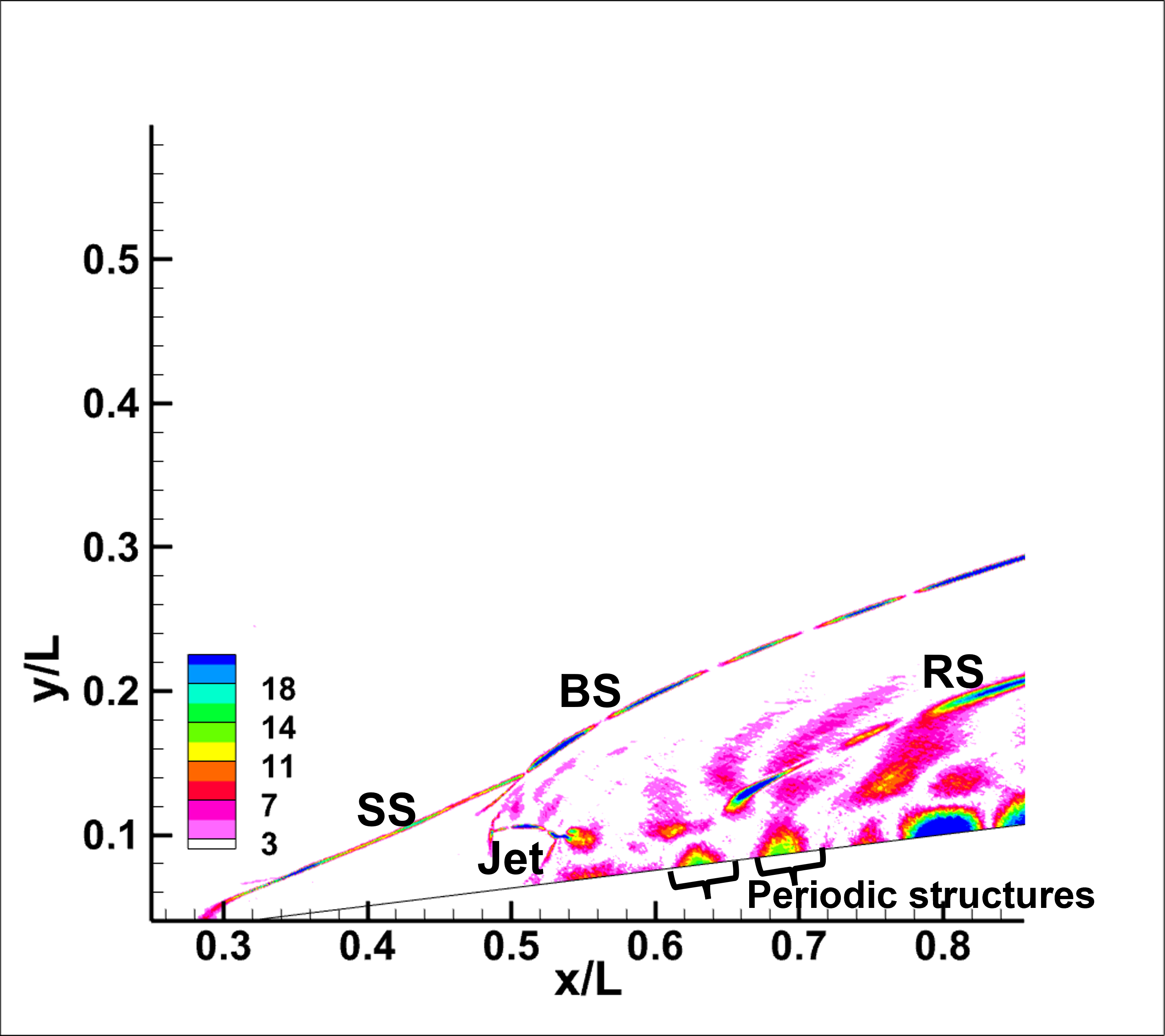}}
\subfigure[ $t_1+3\Delta t_{M6}$ ]{\label{fig:Pressdif2M6}\includegraphics[trim=0.25cm 0.5cm 0.5cm 9cm,clip,width=0.48\linewidth]{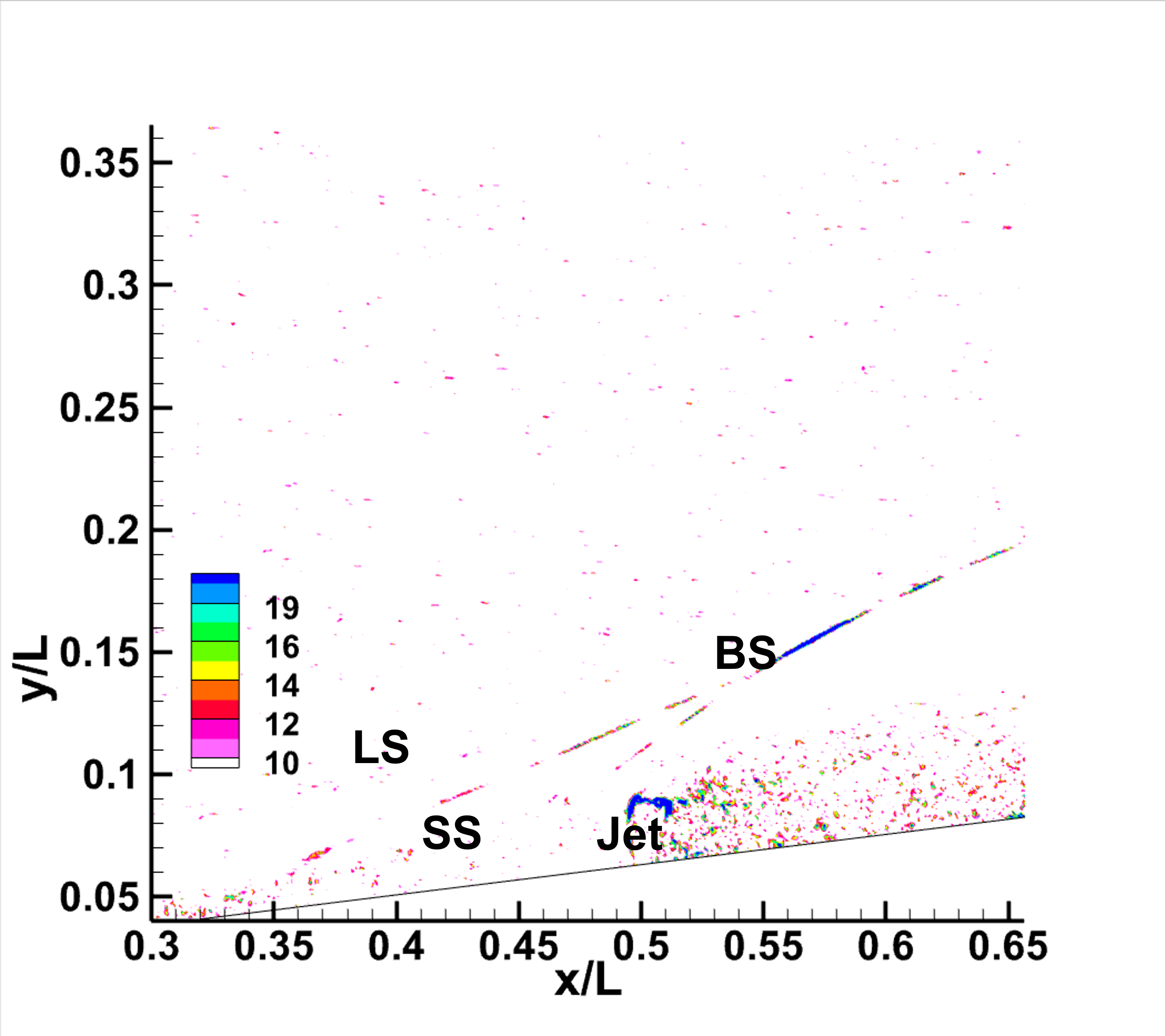}}
\subfigure[ $t_1+3\Delta t_{M12}$ ]{\label{fig:Pressdif3M12}\includegraphics[trim=0.25cm 0.5cm 0.5cm 9cm,clip,width=0.48\linewidth]{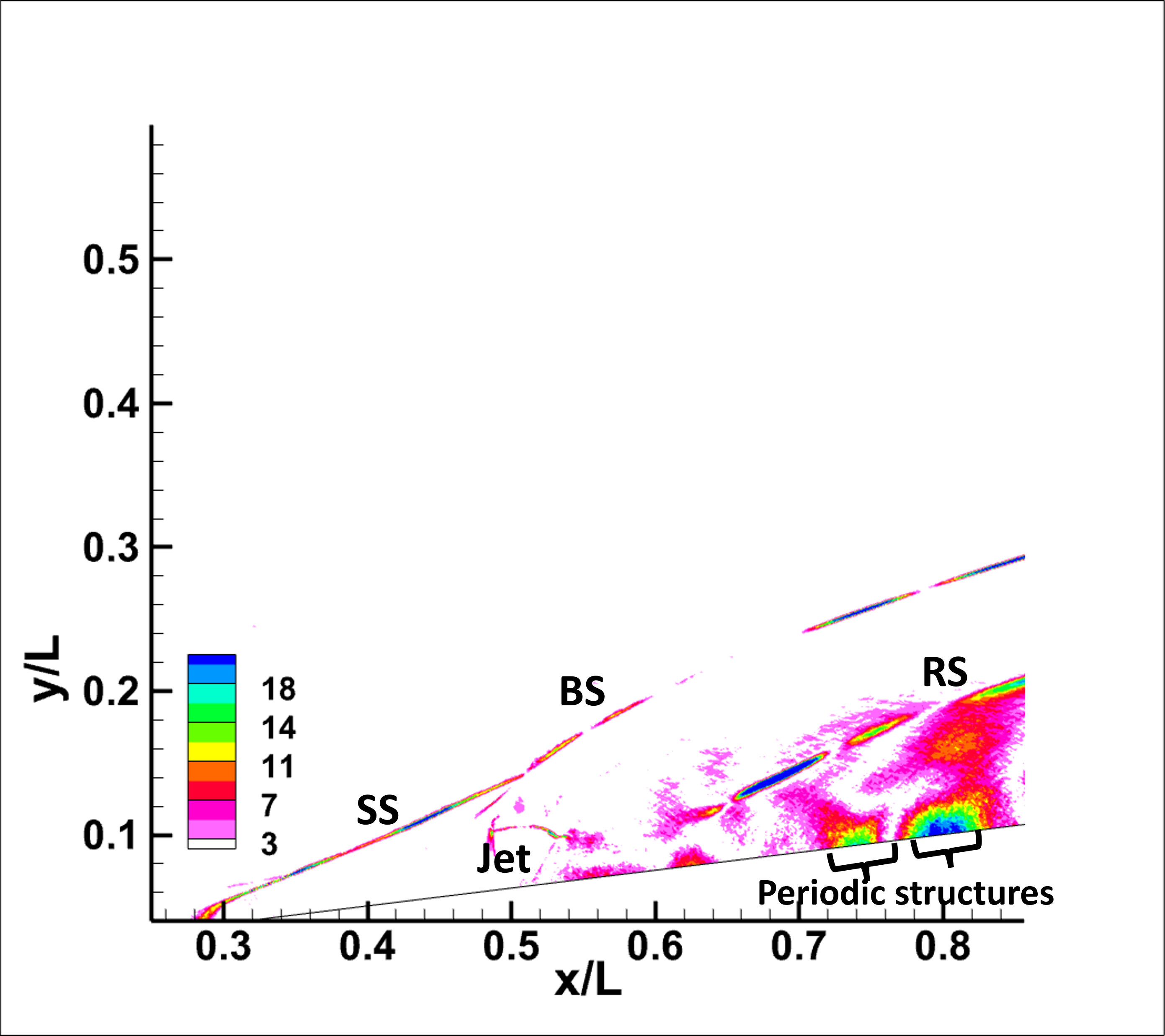}}
\caption{Percent difference in pressure in instantaneous flowfields with indicated shock structures. On the left M6 case, on the right M12 case.}
\label{fig:PercentDiffPress}
\end{center}
\end{figure}

We first consider pressure fluctuations in the flow by calculating the percent difference between two time intervals.  The first time instant, {\em i.e.,} the base flow at $t_1=3 ms$, occurs at a 
 time after the flow establishes which is also an instantaneous flowfield.  The percentage deviation from the base flow at subsequent time intervals, is obtained by subtracting the flowfields at 
 $t_1+\Delta t$, $t_1+2\Delta t$ and $t_1+3\Delta t$ from the base case, 
t$_1$ where $\Delta$t is $0.05$ and $0.015 ms$ for the M6 and M12 cases respectively. Figure~\ref{fig:PercentDiffPress} shows the percent difference in the pressure values for these three time instants for both cases. It can be clearly seen that deviations are on the order of 20\% and higher in the separation-, bow-, and  normal shocks due to the presence of the  jet. It is also observed that in the boundary layer downstream of the jet, fluctuations are even higher than the upstream for both cases with even more structure for the M12 case that resembles a vortex shedding pattern. 

\begin{figure}[htbp]
\begin{center}
\subfigure[Probe locations P1-P6, shown on the Schlieren image for M12 case, but the same interaction locations are used for M6 case.]{\label{fig:ProbeLocs}\includegraphics[trim=0cm 0cm 0cm 9cm,clip,width=0.75\linewidth]{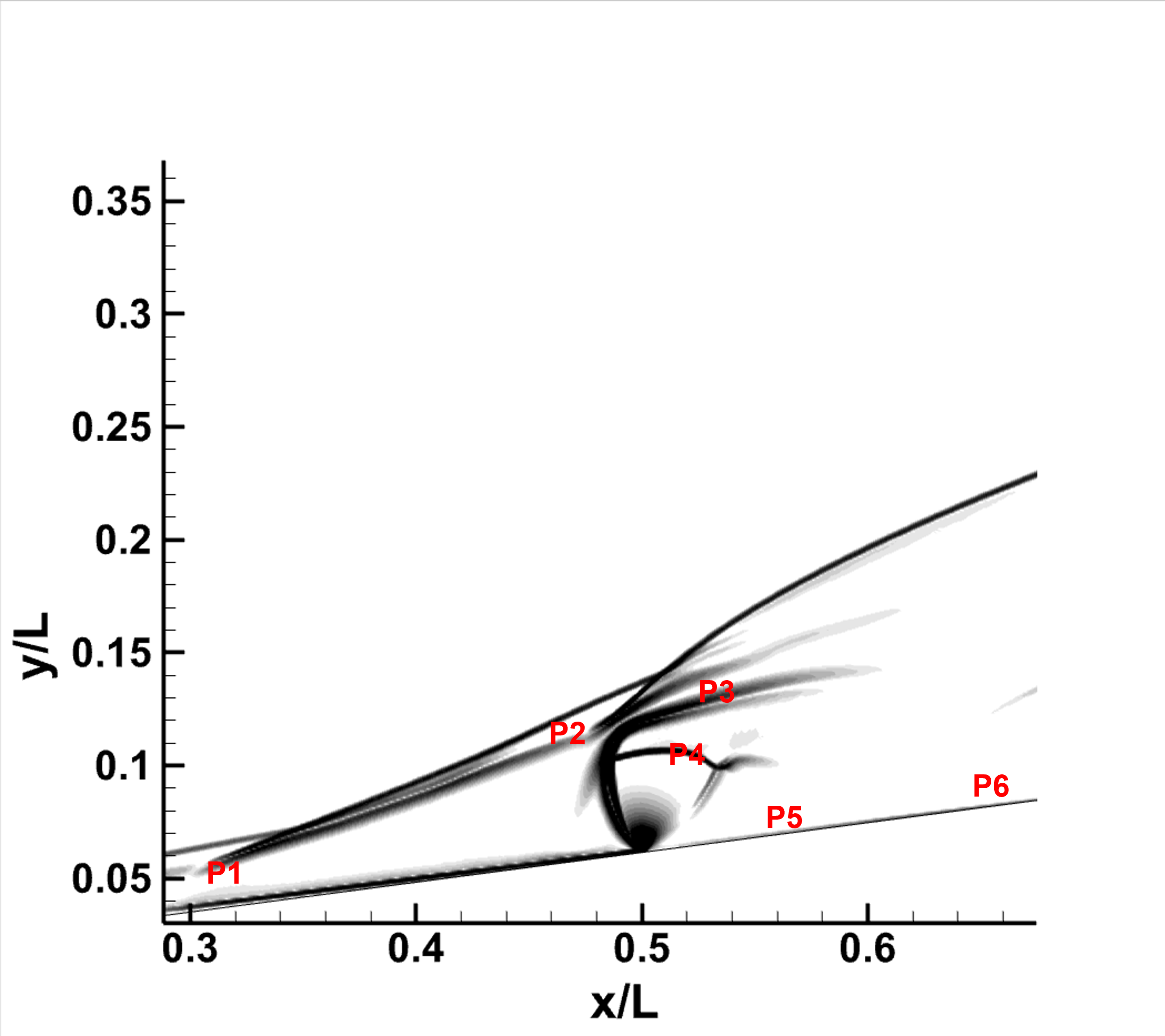}}
\subfigure[Pressure at separation point, M6 case]{\label{fig:PressP6M6}\includegraphics[trim=0cm 0cm 0cm 0cm,clip,width=0.45\linewidth]{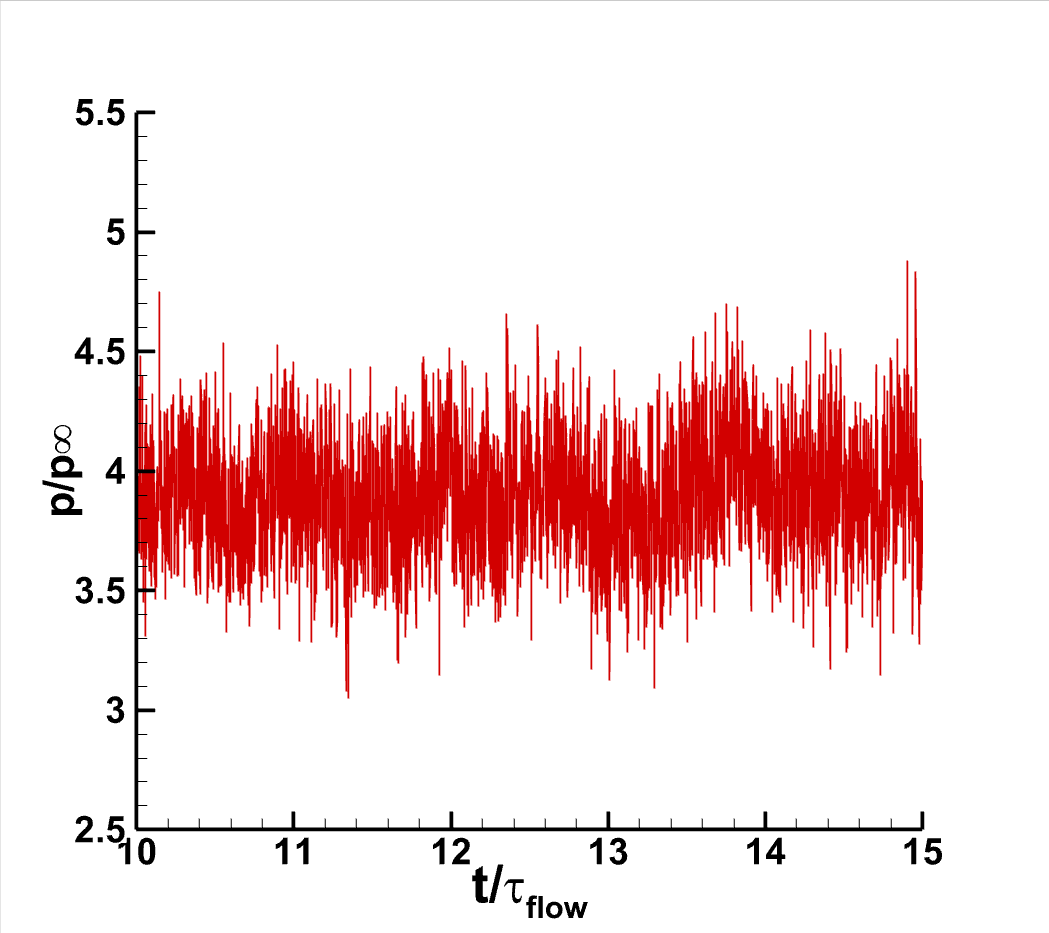}}
\subfigure[PSD of pressure at separation point, M6 case]{\label{fig:PSDP6M6}\includegraphics[trim=0cm 0cm 0cm 0cm,clip,width=0.45\linewidth]{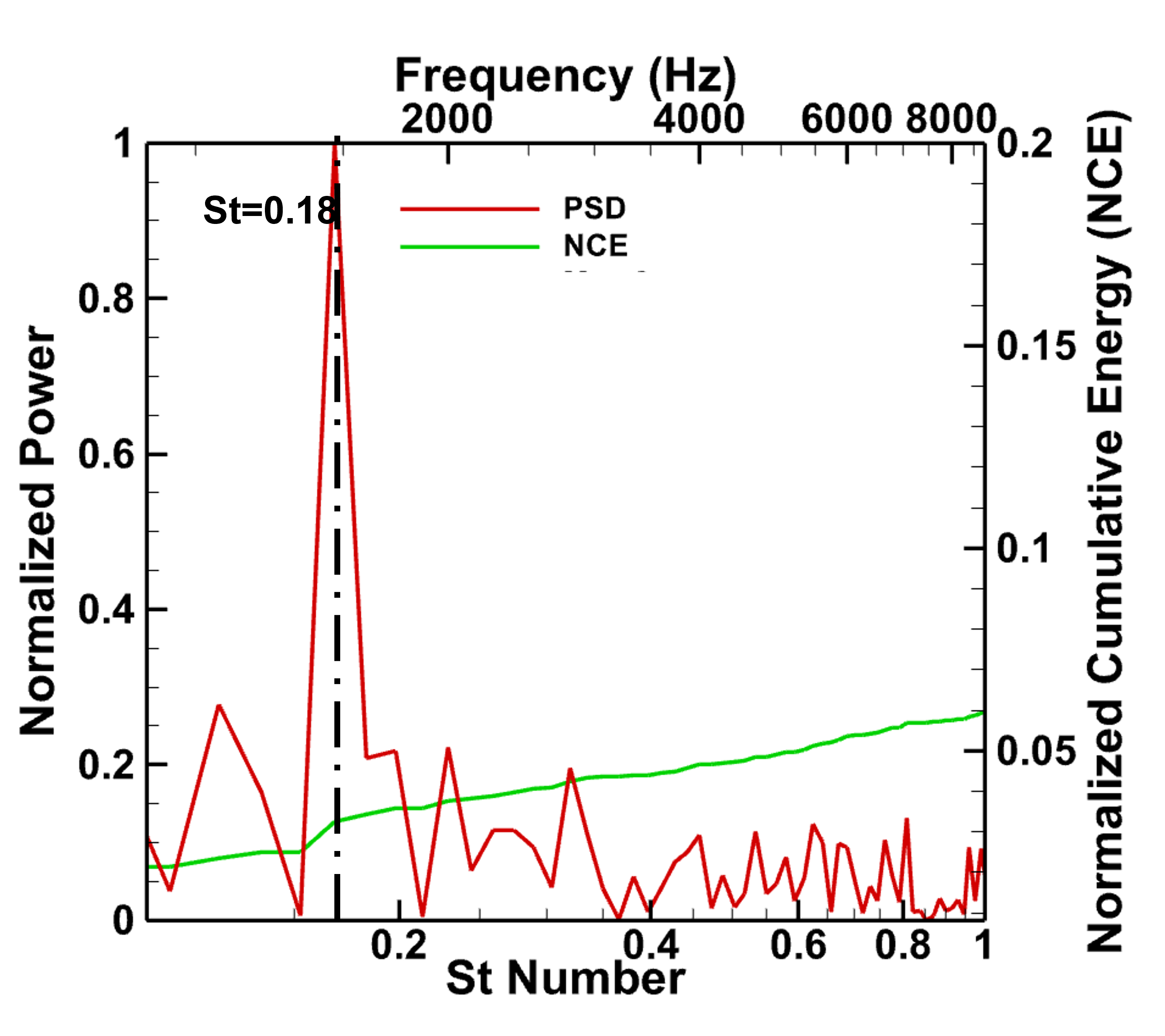}}
\subfigure[Pressure at separation point, M12 case]{\label{fig:PressP2M12}\includegraphics[trim=0cm 0cm 0cm 0cm,clip,width=0.45\linewidth]{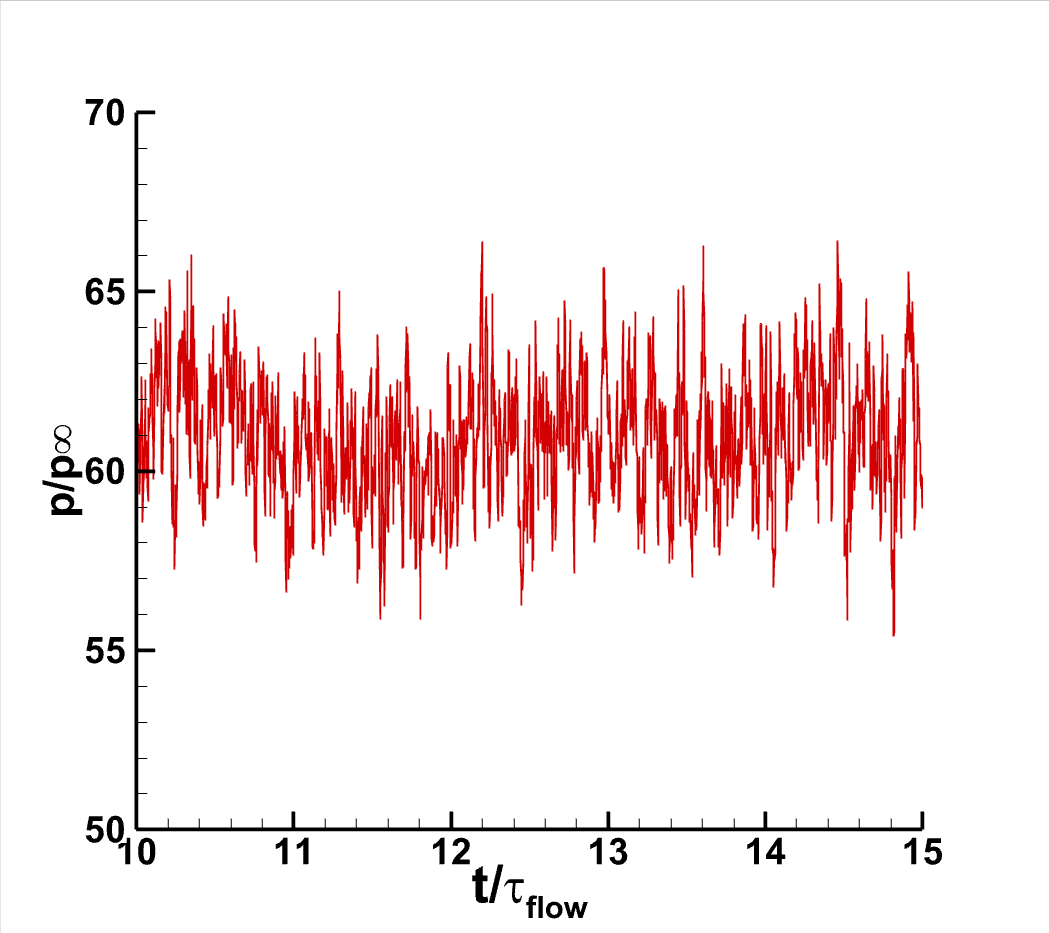}}
\subfigure[PSD of pressure at separation point, M12 case]{\label{fig:PSDP2M12}\includegraphics[trim=0cm 0cm 0cm 0cm,clip,width=0.45\linewidth]{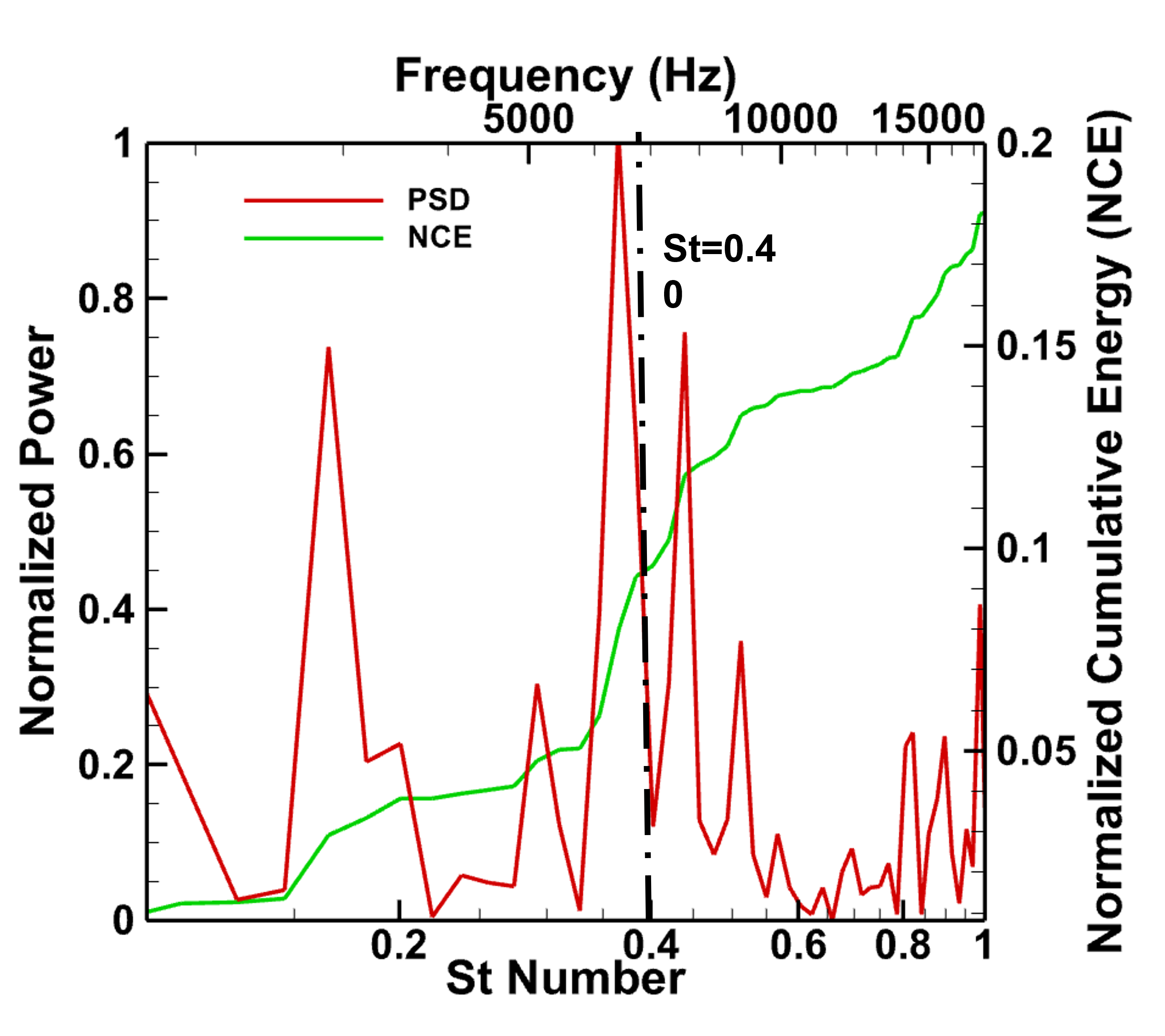}}
\caption{Pressure signal and corresponding Power Spectral Densities (PSD) with normalized cumulative energies (NCE) taken at the separation shock foot, probe location P1.}
\label{fig:ProbePress}
\end{center}
\end{figure}

Next we investigate the detailed time variations of pressure fluctuations using the DSMC data taken from numerical probes place at strategical locations in the flowfield. We place the probes in the locations where shock-shock, shock-boundary layer, shock-shear layer interactions occur and also at locations near the separation-reattachment points, as shown in Fig.~\ref{fig:ProbeLocs}. In the other subfigures of Fig.~\ref{fig:ProbePress} we show the time resolved pressure and its corresponding power spectral density (PSD) plots for the M6 and M12 cases at probe P1  located near the separation shock foot. Similar time behavior was observed for the probes at other interaction regions too but is not shown here. Power spectral densities were calculated for the mean subtracted pressure at these locations. Time resolved DSMC pressure values were collected at a frequency of $200$~MHz, i.e. at each time step of $5ns$, and the sampling was performed for $10 ms$. Then using the Matlab's Fast Fourier Transform algorithm~\cite{MatlabFFT} the  spectral content of the mean subtracted pressure data is extracted. Since DSMC is a particle method, there are always periodic fluctuations due to particle collisions. This is due to two sources, one physical, {\em i.e.,} the thermal speed of the particles, and the second, numerical, {\em i.e.,} the DSMC numerical parameter $FNUM$, the number of real particles each computational particle represents.  The former can never be removed regardless of the number of computational particles whereas the second can be minimized by using as small an FNUM as possible. In Ref.~\cite{sawant1DShock} it was shown that the peaks in the PSD due to DSMC noise decrease as $FNUM$ is decreased, i.e. number of simulation particles are increased, but  other frequencies due to the physical, thermal nature of the flow persist. It can be seen that fluctuations that have lower frequencies {\it lower than} the expected DSMC numerical and thermal fluctuations (order of  MHz) exist in both cases in the current study too, see Figs.~\ref{fig:PSDP6M6} and ~\ref{fig:PSDP2M12}. The power spectral density analysis of the probe DSMC simulation results  show broadband low frequency bands centered around $St=0.19$ and $St=0.4$ for the M6 and M12 cases, respectively where the Strouhal number and the flow time are defined as:

\begin{equation}
\label{eqn:StNumber}
St = \frac{{f{D_{b}}}}{{{U_\infty }}}, {\tau _{flow}} = \frac{L}{{{U_\infty }}}
\end{equation}

\begin{figure}[htbp]
\begin{center}
{\label{fig:StNumber}\includegraphics[trim=0cm 0cm 0cm 0cm,clip,width=0.48\linewidth]{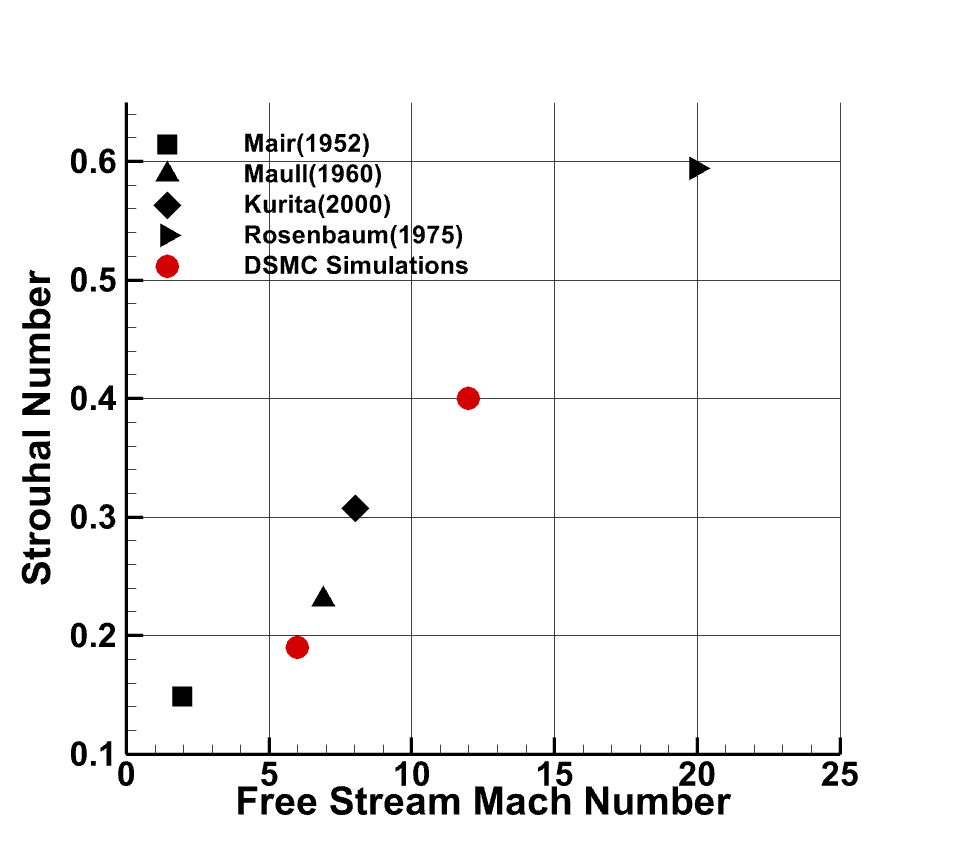}}
\end{center}
\caption{Comparison of St number obtained from this work (red circles) with literature. Note that unit free stream Re numbers (1/m) are on the order of 10$^{7}$ for Mair, Maull and Kurita~\cite{Mair1952,Maull1960,Kurita2000} and 10$^{5}$ for Rosenbaum~\cite{Rosenbaum1975} and this work.}
\label{fig:StNumberComparison}
\end{figure}

These frequencies are near the peak frequencies for each PSD and the inflection point in the normalized cumulative energy (NCE) curves, where the latter is calculated by cumulatively summing up the PSD spectrum. Most importantly, these frequencies are consistent with Strouhal numbers based on cone diameter and  free stream velocity found in the literature for cone geometries, see Fig.~\ref{fig:StNumberComparison}. 
% All the cases shown in Fig.~\ref{fig:StNumberComparison} from the literature are for experiments on cone geometries. 
It appears that the frequency of the unsteadiness in the shock system for the flows with these types of  interaction regions increases with Mach number and does not strongly depend 
on the free stream Reynolds number as long as the incoming boundary layer is laminar.  Moreover, there is evidence that the spectral content of the shock oscillations are insensitive to the free stream Reynolds number but depends on the Mach number for canonical cylinder wake flows too~\cite{roshko_1961,schmidt_shepherd_2015,DuvvuriCylinder2022}.

\begin{figure}[htbp]
\begin{center}
\subfigure[Shear stress, M6 case]{\label{fig:PercentDiffFrictionM6}\includegraphics[trim=0.5cm 0.5cm 0.5cm 0.5cm,clip,width=0.48\linewidth]{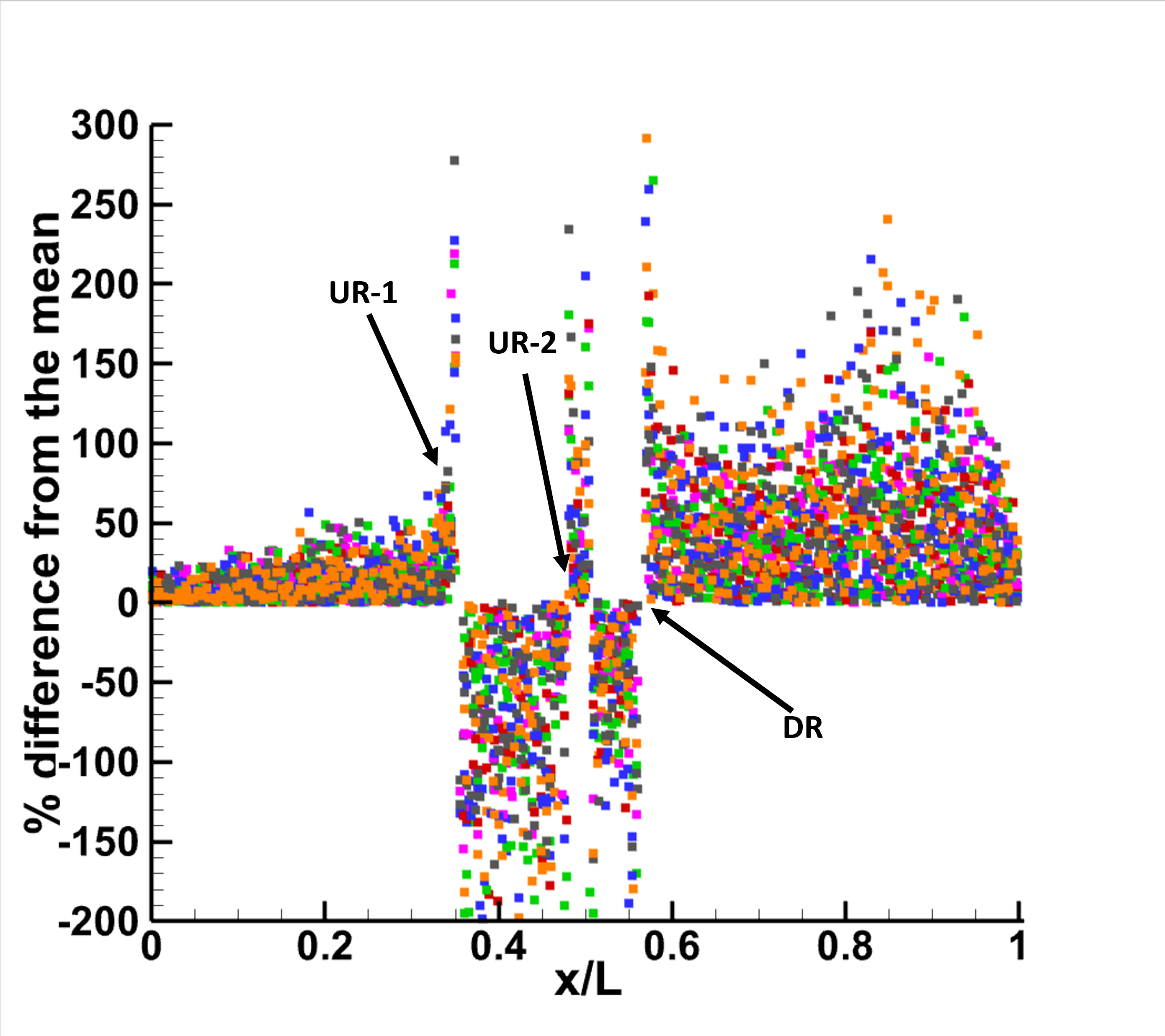}}
\subfigure[Shear stress, M12 case]{\label{fig:PercentDiffFrictionM12}\includegraphics[trim=0.5cm 0.5cm 0.5cm 0.5cm,clip,width=0.48\linewidth]{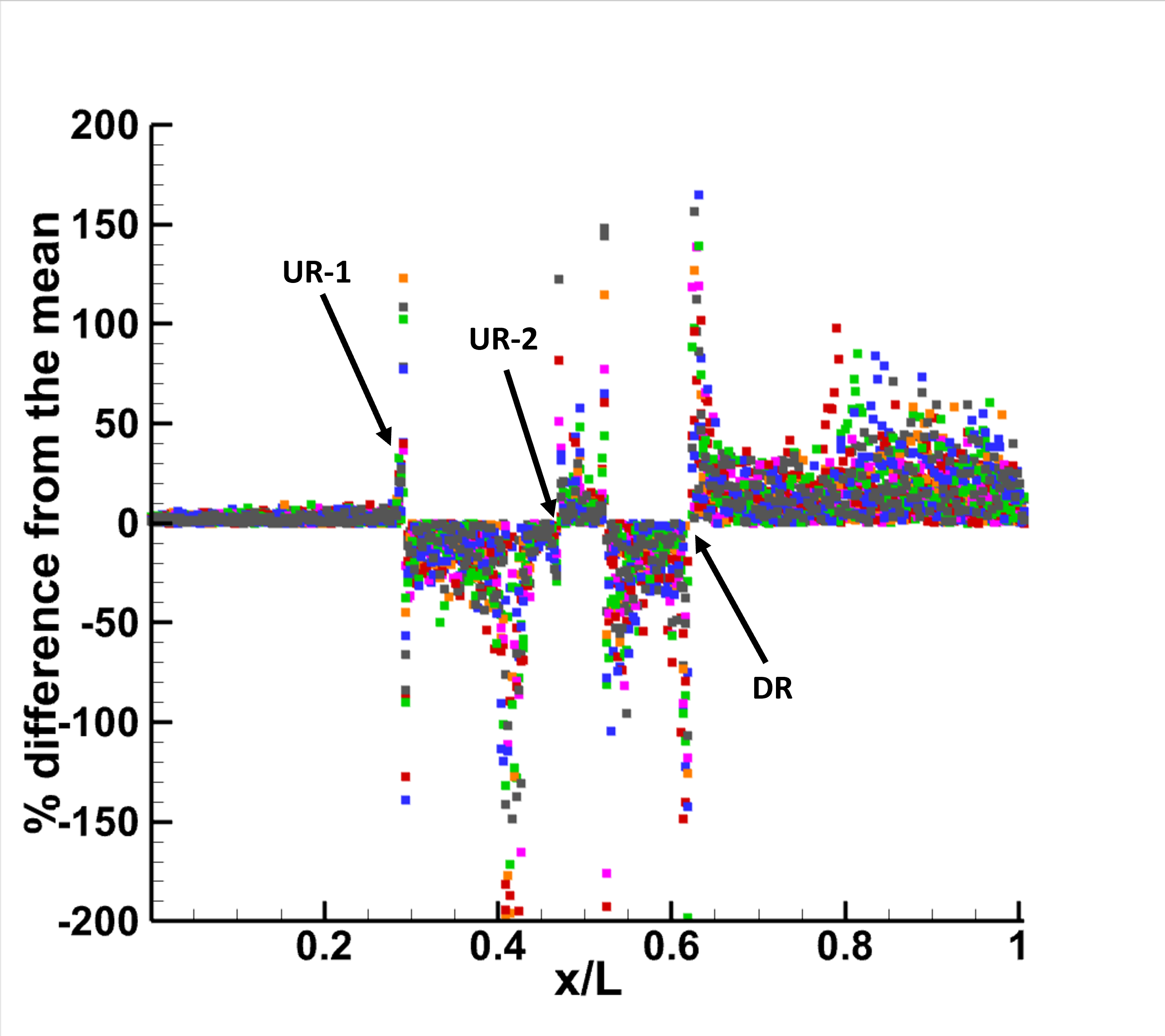}}
\end{center}
\caption{Percent deviation from the mean values of local shear stress. Each color represents a different time instant.}
\label{fig:PercentDiffSurface}
\end{figure}

The low frequency unsteadiness in the flowfield also causes unsteadiness of the local instantaneous surface parameters. To study the effect of this unsteadiness on surface properties, the deviation of the local skin friction coefficient at instantaneous flowfields was calculated at 10 different time intervals for both cases. The first time interval was taken at 10 flow times for both cases and the last one was at 11.4 and 10.7 flow times, respectively for the M6 and M12 cases. The difference between these time instants was chosen as $1/10$-th of the frequencies observed in Figs.~\ref{fig:PSDP6M6} and \ref{fig:PSDP2M12} so that at least one cycle of the shock oscillations was resolved.  Each color in Fig.~\ref{fig:PercentDiffSurface} shows a data point from one of the 10 time intervals.  The skin friction coefficient plots in Figs.~\ref{fig:PercentDiffFrictionM6} and ~\ref{fig:PercentDiffFrictionM12} show 
very high deviations from the mean  especially near the separation and reattachment points. This does not mean that the actual value of the skin friction is vastly different but rather the separation and reattachment locations, where the skin friction is zero, are not stationary. Moreover, the percent deviation from the mean is 50-100\% inside the separation and reattachment regions also, which is much higher than the expected DSMC computational noise (5-10\% for this sampling rate). Another important feature of the plots in Fig.~\ref{fig:PercentDiffSurface} is that the deviation from the mean is much higher in the downstream boundary layer compared to the upstream boundary layer consistent with Fig.~\ref{fig:TTrnFracMixing} which showed that the downstream boundary layer consists of purely the jet gas with vortex shedding like motion due to the jet expansion, as observed in Figs.~\ref{fig:Pressdif1M12},~\ref{fig:Pressdif2M12} and ~\ref{fig:Pressdif3M12}. Both of these flow features indicate high unsteadiness.

\section{Conclusions}\label{sec:Conc}

The interactions between a side jet and hypersonic flows at M=6 and M=12 over a slender cone  were simulated using the DSMC method. Both the time averaged steady state and time resolved results of shock structures and surface parameters were examined. It was demonstrated that continuum breakdown occurs in the leading edge shock, bow shock and especially in the jet expansion which motivates the use of a kinetic method. Steady state results showed that the presence of the jet on the cone increases the complexity of the  flowfield by creating a complex shock structure, recirculation regions and shear layers.   We demonstrated that many of the key normalized parameters, such as penetration height and shock angles, are captured well by the DSMC simulations and 
agree  with theoretical and experimental results. 
It was shown that the jet and free stream gases mix only in the upstream separation region and the boundary layer flow downstream of the jet consists only of the jet gas, resulting in negative heat transfer values downstream of the jet. Integral values for the surface parameters showed that drag force and heat transfer decrease and force in the direction normal to the free stream increases with the presence of the jet for both Mach number cases. Instantaneous flowfield DSMC pressures showed high pressure fluctuations, especially around the shocks and jet expansions with a low frequency unsteadiness consistent with the literature 
that can be attributed to these fluctuations.  These fluctuations in the shocks and boundary layer also affects the surface parameters, especially drag force, where deviations on the order of 100\% were observed in the jet-interaction region.

\section*{Acknowledgments}
Computational resources for this research is provided by the National Science Foundation TACC's Stampede2 supercomputer and Purdue University's Anvil supercomputer, which is a part of ACCESS (previously XSEDE), with project number TG-PHY210065. Authors are grateful to Dr. Joseph Jewell from Purdue University for providing the case setup and extensive discussions on the subject.
\pagebreak
\bibliography{JetRefs}
\pagebreak
\pagenumbering{arabic}
\end{document}